\documentclass[12pt]{article}
\usepackage{amssymb,amsmath,epsfig}
\usepackage{graphicx}
\usepackage{amsfonts,graphicx,amssymb,amsmath,amsthm,epsfig}
\usepackage{float}
\allowdisplaybreaks
\begin{document}
\title{\bf Analysis of Reconstructed Modified $f(Q,T)$ Gravity}
\author{M. Sharif \thanks {msharif.math@pu.edu.pk}~and Iqra Ibrar
\thanks{iqraibrar26@gmail.com}\\
Department of Mathematics and Statistics, The University of Lahore,\\
1-KM Defence Road Lahore-54000, Pakistan.}
\date{}
\maketitle

\begin{abstract}
This paper studies the reconstruction criterion in the framework of
$f(Q,T)$ gravity using ghost dark energy model, where $Q$ represents
the non-metricity and $T$ is the trace of energy-momentum tensor. In
this regard, the correspondence scenario for a non-interacting
scheme is used to construct the ghost dark energy $f(Q,T)$ model. We
consider the flat Friedmann-Robertson-Walker universe with power-law
scale factor and pressure-less matter. We investigate the behavior
of equation of state parameter and examine the stability of ghost
dark energy model via squared speed of sound parameter. The equation
of state parameter depicts the phantom era, while the squared speed
of sound reveals stable ghost dark energy model for the entire
cosmic evolutionary paradigm. Finally, we study the cosmography of
the $\omega_{GDE}-\omega'_{GDE}$ and $r-s$ planes that correspond to
freezing region and Chaplygin gas model, respectively. It is
concluded that the reconstructed ghost dark energy model represents
the evolution of the cosmos for suitable choices of the parametric
values.
\end{abstract}

\section{Introduction}

The cosmological and astrophysical phenomena happening in our
universe have inspired numerous scholars to explore its enigmatic
aspects. The universe is comprised of three distinct components,
i.e., dark energy (DE), dark matter (DM) and ordinary source. The
cosmos is largely dominated by DE and DM, while the remaining
portion is occupied by the usual matter. Dark matter is an invisible
and non-baryonic source whose existence has been clarified by some
phenomena such as gravitational lensing and galactic rotation curves
\cite{1}. On the other hand, the most ground-breaking investigation
of the last few decades was the accelerated expanding behavior of
the cosmos, which provoked scientist's interest in an entirely novel
direction. An unknown energy source with a large negative pressure
known as DE is believed to be the origin of this expansion. The
problems associated with the nature and occurrence of DE and DM are
considered among the most challenging and unresolved issues in
cosmology. The best model so far to elucidate the nature of DE is
the standard $\Lambda$CDM paradigm. The cosmological constant is
supposed to be the best candidate to signify the accelerated
phenomenon of the cosmos. Despite of its consistent behavior with
the observational data, it faces some challenges like fine tuning
and cosmic coincidence \cite{3}. There are basically two ways to
unravel the mysterious origin of cosmic accelerated expansion. One
strategy is to reshape the geometric part of the Einstein-Hilbert
action which leads to modified theories of gravity while the other
possibility lies in changing the matter portion into dynamic DE
models \cite{4}.

Scholars have provided several approaches to tackle these issues in
recent years, but they remain mysterious to this day. In order to
describe the nature of DE, the equation of state (EoS) parameter
($\omega=\frac{P}{\rho},~\rho$ and $P$ represent the energy density
and pressure, respectively) is found to be helpful for the model
under consideration. Recently, the Veneziano ghost DE (GDE) has been
suggested \cite{5} which is non-physical in the standard Minkowski
spacetime but has significant physical consequences in non-trivial
topological or dynamical spacetimes. In curved spacetime, it
produces a small vacuum energy density that is proportional to
$\Lambda^{3}_{QCD}H$, where $\Lambda_{QCD}$ is the quantum
chromodynamics $(QCD)$ mass scale and $H$ is the Hubble parameter.
Thus, no further parameters, degrees of freedom, or modifications
are required. With $\Lambda_{QCD}\sim 100 MeV$ and $H\sim
10^{-33}eV$, $\Lambda^{3}_{QCD}H$ gives the right order of observed
DE density. This remarkable coincidence suggests that the
fine-tuning issues are removed by this model \cite{6}.

The well-known gravitational theory known as GR is modified to
several gravitational theories which attracted many people due to
their beneficial features in defining the accelerated expansion of
the cosmos \cite{21}. According to the Weyl theory, the covariant
divergence of the metric tensor is non-zero. This characteristic can
be mathematically expressed in terms of a recently discovered
non-metricity quantity. Dirac \cite{21-a} suggested a generalization
of Weyl theory based on the idea that there are basically two
metrics, however, this notion was generally ignored by physicists. A
new generalized geometric theory called Einstein-Cartan theory
\cite{21-b} was developed as a result of significant contributions
of Cartan to geometric theories. The resulting geometry is known as
the Weyl-Cartan (WC) geometry \cite{21-d}. Another modification was
introduced by Weitzenb$\ddot{o}$ck in which the metric, curvature
and torsion tensors, respectively, of a Weitzenb$\ddot{o}$ck
manifold are represented by the properties
$\nabla_{\zeta}g_{\eta\lambda}=0$,
$\mathcal{W}^{\zeta}_{\eta\lambda\sigma}=0$ and
$T^{\zeta}_{\eta\lambda}=0$ \cite{21-e}. Moreover, the condition
$T^{\zeta}_{\eta\lambda}=0$ transforms the Weitzenb$\ddot{o}$ck
space into a Euclidean manifold. The Riemann curvature tensor of
Weitzenb$\ddot{o}$ck space is zero, due to which these geometries
possess significant feature of distant parallelism, also known as
tele parallelism or absolute parallelism.

The teleparallel equivalent to GR is an alternative geometrical
description of GR. The basic principle behind teleparallel gravity
is to replace the metric $g_{\alpha\beta}$ (the fundamental physical
variables) with the collection of tetrad vectors $e^{i}_{\alpha}$,
which in turn yields torsion. Moreover, the curvature is then
replaced by the torsion generated by the tetrad, which can be used
to describe the gravitational effects of the universe \cite{23}.
Linder \cite{18} proposed the so called modified teleparallel
gravity, also known as $f(\mathbb{T})$ gravity, where $\mathbb{T}$
represents the torsion scalar. Meanwhile, the present cosmic rapid
expansion is widely explained by $f(\mathbb{T})$ gravity, which does
not depend on the notion of DE \cite{24}. There can be two
equivalent geometric representations of GR among which one is the
curvature representation in which the non-metricity and torsion
vanish. The second case gives rise to the teleparallel formalism, in
which the role of non-metricity as well as the curvature disappear.
A third equivalent representation is also possible, where the
gravitational interaction is represented by the non-metricity $Q$ of
the metric signifies the change in vector length during parallel
transport. In 1999, for the very first time, Nester and Yo
introduced the concept of symmetric teleparallel gravity (STG)
\cite{13}.

The STG was extended to $f(Q)$ theory, introduced by Jimenez et al.
\cite{41}, also named as non-metric gravity. Using this theory, it
has been revealed that the rapid expansion of the cosmos is
intrinsic characteristic and does not require any exotic DE or
additional field \cite{25}. Lazkoz et al. \cite{7} described the
constraints on $f(Q)$ gravity by using polynomial functions of the
red-shift. The energy conditions for two distinct $f(Q)$ gravity
models have also been determined \cite{31}. The behavior of cosmic
parameters in the same context was examined by Koussour et al.
\cite{9}. Chanda and Paul \cite{11} studied the formation of
primordial black holes in this theory.

Recently, Xu et al. \cite{20} extended this theory to $f(Q,T)$
gravity by coupling the non-metricity with the trace of
energy-momentum tensor. The Lagrangian of the gravitational field is
assumed to be a general function of both $Q$ and $T$. The field
equations of the respective theory are obtained by varying the
gravitational action with respect to both metric and connection. The
purpose to present this theory is to examine its theoretical
consequences, consistency with actual experimental evidences and
adaptability to cosmological scenarios. The cosmological evolution
equations for a flat homogeneous isotropic geometry are obtained by
assuming a simple functional form of this theory. Bhattacharjee and
Sahoo \cite{44} examined the gravitational baryogenesis interactions
in the same theory and concluded that the respective phenomenon is
found consistent. Arora et al. \cite{45} studied the accelerated
expansion of the universe without introducing additional forms of DE
in this gravity theory. Pati et al. \cite{46} investigated the
dynamical features and cosmic evolution in the corresponding
gravity. Xu et al. \cite{47} generalized the Friedman equations and
investigated the cosmological implications by assuming some
particular functional forms.

The accelerated expansion of the cosmos can be analyzed by several
kinds of DE models. Turner and White \cite{42} found that the
inflationary era contradicted with the current matter density and
resolved this dilemma by parameterizing the smooth component. Sahni
et al. \cite{33} proposed a dimensionless diagnostic pair namely the
state finder parameters to examine the nature of DE through EoS.
Sharif and Zubair \cite{17} reconstructed the holographic DE model
to evaluate the phantom and non-phantom phases for the cosmic
expanding regime. Taking an ideal fluid source, Chirde and Shekh
\cite{12} studied a non-static plane symmetric DE model with the
help of EoS and skewness parameter. Saba and Sharif \cite{16}
reconstructed the cosmic behavior of agegraphic DE $f(G,T)$ ($G$
indicates the Gauss-Bonnet term) models through diagnostic
parameters and phase planes. The same authors \cite{54} examined the
cosmic diagnostic parameters and reconstructed the $f(G,T)$ gravity
using the Tsallis holographic DE. Solanki et al. \cite{14} analyzed
the profile of different cosmological parameters in DE $f(Q)$
gravity with the help of energy conditions. Mussatayeva et al.
\cite{15} explained a variety of late-time cosmic occurrences in
$f(Q)$ theory. Arora et al. \cite{71} discussed the late-time
cosmology undergoing pressure-less matter by choosing a specific
$f(Q,T)$ model.

Reconstruction phenomena in modified theories of gravity provide a
helpful source to construct a realistic GDE model that describes the
cosmic evolutionary history. The associated energy densities of the
modified theories and the DE model are compared in the
reconstruction scenario. Cai et al. \cite{37} studied the
cosmological evolution of DE interaction with CDM model. Sheykhi and
Movahed \cite{28} observed the expansion of the cosmos through model
parameter constraints in GR for an interacting GDE paradigm. The
role of thermodynamics to understand the cosmic accelerated
expansion through GDE model has been discussed in \cite{38}. Saaidi
et al. \cite{72} investigated the density parameter, squared
adiabatic sound speed, EoS and deceleration parameters by studying
the correspondence between $f(R)$ ($R$ is the Ricci scalar) and GDE
model. Sharif and Saba \cite{52} reconstructed GDE in the context of
$f(G)$ gravity for both interacting and non-interacting schemes to
explain the evolution of the universe. Myrzakulov et al. \cite{51}
investigated the evolution trajectories of the EoS parameter,
$\omega_{D}-\omega'_{D}$ and the state finder planes in ghost and
pilgrim DE $f(Q)$ gravity. Reconstruction of $f(Q)$ theory on the
basis of several distinct models is examined to study the evolution
of the universe \cite{10}.

In this paper, we adopt the correspondence technique to reconstruct
GDE $f(Q,T)$ model and discuss some significant features. The paper
has the following format. In the next section, we present a detailed
discussion of $f(Q,T)$ gravity. We investigate the reconstruction
procedure to formulate the GDE $f(Q,T)$ paradigm in section
\textbf{3}. We figure out the cosmic behavior through EoS parameter,
$\omega_{GDE}-\omega'_{GDE}$ and $r-s$ planes in section \textbf{4}.
We also study the stability of this model through squared speed
component. The final section presents our main results.

\section{Basic Formalism of $f(Q,T)$ Theory}

In this section, the basic structure of modified $f(Q,T)$ theory is
presented using the variational principle in order to formulate the
field equations. Weyl \cite{60} introduced a generalization of
Riemannian geometry as a mathematical framework for describing
gravitation in GR. In Riemannian geometry, parallel transport around
a closed path preserves direction and length of a vector. This
adjustment includes a new vector field $\omega^{\alpha}$ that
characterizes the geometric features of Weyl geometry. The vector
field is introduced to account for a modification in length during
parallel transport while the metric tensor maintains the local
structure of spacetime by specifying distances and angles. According
to Weyl's theory, a vector field has similar mathematical properties
as electromagnetic potentials in physics, which denotes a strong
connection between gravitational and electromagnetic forces. Both of
these forces are considered as long-range forces and Weyl's proposal
raises the possibility of a common geometric origin for these forces
\cite{21-a}.

A vector length $y$ in a Weyl geometry transports if it is moved
along an infinitesimal path $\delta x^{\alpha}$, meaning that its
new length is represented by $\delta y = y \omega_{\alpha }\delta
x^{\alpha}$ \cite{21-a}. After the parallel transport of a vector
around a small closed loop of area $\delta s^{\alpha\beta}$, the
variation in the length of a vector is given by the expression
$\delta y=y\chi_{\alpha\beta}\delta s^{\alpha\beta}$ with
\begin{equation}\label{1}
\chi_{\alpha\beta}=\nabla_{\beta}\omega_{\alpha}-\nabla_{\alpha}\omega_{\beta}.
\end{equation}
This indicates that the area enclosed by the loop, the curvature of
the Weyl connection and the original length of the vector are all
proportional to the variation in the vector's length. A local
scaling of lengths of the form $\tilde{y}=\phi(x)y$ changes the
field $\omega_{\alpha}$  to
$\tilde\omega_{\alpha}=\omega_{\alpha}+(\ln\phi),_{\alpha}$ in which
the metric coefficients are modified as
$\tilde{g}_{\alpha\beta}=\omega^{2}g_{\alpha\beta}$ and
$\tilde{g}^{\alpha\beta}=\omega^{-2}g^{\alpha\beta}$ under the
conformal transformations \cite{21-d}. The existence of semi-metric
connection is another vital component of the Weyl geometry given as
\begin{equation}\label{2}
\bar{\Gamma}^{\lambda}_{\alpha\beta}=\Gamma^{\lambda}_{\alpha\beta}
+g_{\alpha\beta}\omega^{\lambda}-\delta^{\lambda}_{\alpha}
\omega_{\beta}-\delta^{\lambda}_{\beta}\omega_{\alpha},
\end{equation}
where $\Gamma^{\lambda}_{\alpha\beta}$ represents the usual
Christoffel symbol. One can construct a gauge covariant derivative
based on the supposition that $\bar{\Gamma}^{\lambda}_{\alpha\beta}$
is symmetric \cite{21-d}. The Weyl curvature tensor can be obtained
using the covariant derivative which can be written as
\begin{equation}\label{3}
\mathcal{\bar{W}}_{\alpha\beta\rho\mu}=
\mathcal{\bar{W}}_{(\alpha\beta)\rho\mu}+\mathcal{\bar{W}}
_{[\alpha\beta]\rho\mu},
\end{equation}
where
\begin{align*}
\mathcal{\bar{W}}_{[\alpha\beta]\rho\mu}&=\mathcal{W}_{\alpha\beta\rho\mu}
+2\nabla_{\rho}\omega_{[\alpha g_{\beta}]\mu} +
2\nabla_{\mu}\omega_{[\beta g_{\alpha}]\rho}
+2\omega_{\rho}\omega_{[\alpha
g_{\beta}]\mu}+2\omega_{\mu}\omega_{[\beta g_{\alpha}]\rho}
\\
& -2\omega^{2}g_{\rho{[\alpha g_{\beta}]}\mu},
\\
\mathcal{\bar{W}}_{(\alpha\beta)\rho\mu}&=\frac{1}{2}(\mathcal{\bar{W}}_{\alpha\beta\rho\mu}
+\mathcal{\bar{W}}_{\beta\alpha\rho\mu})=g_{\alpha\beta}\chi_{\rho\mu}.
\end{align*}
The Weyl curvature tensor after the first contraction yields
\begin{equation}\label{6}
\mathcal{\bar{W}}^{\alpha}_{\beta}=\mathcal{\bar{W}}^{\rho\alpha}_{\rho\beta}
=\mathcal{W}^{\alpha}_{\beta}+2\omega^{\alpha}\omega_{\beta}
+3\nabla_{\beta}\omega^{\alpha} -\nabla_{\alpha}\omega^{\beta}
+g^{\alpha}_{\beta}(\nabla_{\rho}\omega^{\rho}-2\omega_{\rho}\omega^{\rho}).
\end{equation}
Its contraction yields the Weyl scalar as
\begin{equation}\label{7}
\mathcal{\bar{W}}=\mathcal{\bar{W}}^{\rho}_{\rho}=
\mathcal{W}+6(\nabla_{\alpha}\omega^{\alpha}-\omega_{\alpha}\omega^{\alpha}).
\end{equation}

In addition to Riemannian and Weyl geometries, WC spaces with
torsion represent a more general framework. In WC spacetime, the
length of a vector is defined by the metric tensor and parallel
transport law is determined by a symmetric connection
$d\vartheta^{\alpha}=-\vartheta^{\phi}\tilde{\Gamma}^{\alpha}_{\phi\beta}dx^{\beta}$
\cite{21-b}. The connection for the WC space is expressed as
\begin{equation}\label{8}
{\hat\Gamma^{\lambda}_{\alpha\beta}}=\Gamma^{\lambda}_{\alpha\beta}
+\mathcal{C}^{\lambda}_{\alpha\beta}+\mathcal{L}^{\lambda}_{\alpha\beta},
\end{equation}
where $\mathcal{C}^{\lambda}_{\alpha\beta}$ and
$\mathcal{L}^{\lambda}_{\alpha\beta}$ indicate the contortion tensor
and the disformation tensor, respectively. The contorsion tensor
from the torsion tensor can be derived as
\begin{equation}\label{11}
\mathcal{C}^{\lambda}_{\alpha\beta}=\hat\Gamma^{\lambda}_{[\alpha\beta]}
+g^{\lambda\phi}g_{\alpha\kappa}\hat\Gamma^{\kappa}_{[\beta\phi]}
+g^{\lambda\phi}g_{\beta\kappa}\hat\Gamma^{\kappa}_{[\alpha\phi]}.
\end{equation}
From the non-metricity, the disformation tensor can be obtained as
\begin{equation}\label{12}
\mathcal{L}^{\lambda}_{\alpha\beta}=\frac{1}{2}g^{\lambda\phi}(Q_{\beta\alpha\phi}
+Q_{\alpha\beta\phi}-Q_{\lambda\alpha\beta}),
\end{equation}
where
\begin{equation}\label{13}
Q_{\lambda
\alpha\beta}=\nabla_{\lambda}g_{\alpha\beta}=-g_{\alpha\beta},_{\lambda}
+g_{\beta\phi}\hat{\Gamma}^{\phi}_{\alpha\lambda}
+g_{\phi\alpha}\hat{\Gamma}^{\phi}_{\beta\lambda}.
\end{equation}
It is clearly shown in Eqs.\eqref{2} and \eqref{8} that the WC
geometry with zero torsion is a particular form of the Weyl geometry
and the non-metricity is determined as $Q_{\lambda\alpha\beta} =
-2g_{\alpha\beta}w_{\lambda}$. Consequently, Eq.\eqref{8} becomes
\begin{equation}\label{14}
\hat{\Gamma}^{\lambda}_{\alpha\beta}={\Gamma}^{\lambda}_{\alpha\beta}
+g_{\alpha\beta}w^{\lambda}-\delta^{\lambda}_{\alpha}w_{\beta}-
\delta^{\lambda}_{\beta}w_{\alpha} +C^{\lambda}_{\alpha\beta},
\end{equation}
with
\begin{equation}\label{15}
C^{\lambda}_{\alpha\beta}=T^{\lambda}_{\alpha\beta}-g^{\lambda\mu}
g_{\phi\alpha}T^{\phi}_{\mu\beta}-g^{\lambda\mu}g_{\phi\beta}
T^{\phi}_{\mu\alpha},
\end{equation}
and the WC torsion is defined as
\begin{equation}\label{16}
T^{\lambda}_{\alpha\beta}=\frac{1}{2}(\hat{\Gamma}^{\lambda}_{\alpha\beta}
-\hat{\Gamma}^{\lambda}_{\alpha\beta}).
\end{equation}
Using the connection, the WC curvature tensor is described by
\begin{equation}\label{17}
\hat{W}^{\lambda}_{\alpha\beta\phi}=\hat{\Gamma}^{\lambda}_{\alpha\phi,\beta}
-\hat{\Gamma}^{\lambda}_{\alpha\beta,\phi}+\hat{\Gamma}^{\rho}_{\alpha\phi}
\hat{\Gamma}^{\lambda}_{\alpha\beta}-\hat{\Gamma}^{\rho}_{\alpha\beta}
\hat{\Gamma}^{\lambda}_{\rho\phi}.
\end{equation}
The WC scalar can be achieved by contracting the curvature tensor as
\begin{eqnarray}\nonumber
\hat{W}&=&\hat{W}^{\alpha\beta}_{\alpha\beta}
=W+6\nabla_{\beta}\omega^{\beta}-4\nabla_{\beta}T^{\beta}
-6\omega_{\beta}\omega^{\beta}+8\omega_{\beta}T^{\beta} +T^{\alpha
\rho\beta}T_{\alpha\rho\beta}
\\\label{18}
&+&2T^{\alpha\rho\beta}T_{\beta\rho\alpha}-4T_{\beta}T^{\beta},
\end{eqnarray}
where $T_{\alpha}= T^{\beta}_{\alpha\beta}$ and all the covariant
derivatives are taken with respect to the metric tensor. The
connection in terms of disformation tensor is expressed as
\begin{align}\label{19}
\Gamma^{\lambda}_{\alpha\beta}&=-\mathcal{L}^{\lambda}_{\alpha\beta}.
\end{align}

The gravitational action in a non-covariant form \cite{27} is given
as
\begin{equation}\label{20}
S=\frac{1}{2k} \int g^{\alpha\beta}(\Gamma^{\rho}_{\phi\alpha}
\Gamma^{\phi}_{\beta\rho} -\Gamma^{\rho}_{\phi\rho}
\Gamma^{\phi}_{\alpha\beta})\sqrt{-g}d^{4}x.
\end{equation}
Using the relation \eqref{19}, the action integral takes the form
\begin{equation}\label{21}
S=-\frac{1}{2k} \int g^{\alpha\beta}(\mathcal{L}^{\rho}_{\phi\alpha}
\mathcal{L}^{\phi}_{\beta\rho} - \mathcal{L}^{\rho}_{\phi\rho}
\mathcal{L}^{\phi}_{\alpha\beta}) \sqrt{-g} d^{4}x.
\end{equation}
This action is known as the action of STG which is equivalent to the
Einstein-Hilbert action. There are some significant distinctions
between two gravitational paradigms. One of them is that the
vanishing of curvature tensor in STG causes the system to appear as
flat structure throughout. Furthermore, the gravitational effects
occur due to variations in the length of vector itself, rather than
rotation of an angle formed by two vectors in parallel transport.

Now, we look at an extension of STG Lagrangian stated as
\begin{equation}\label{22}
S=\int\bigg[\frac{1}{2k}f(Q,T)+L_{m}\bigg]\sqrt{-g}d^{4}x,
\end{equation}
$g$ indicates the determinant of the metric tensor, $L_{m}$
represents the matter Lagrangian and $k$ signifies the coupling
constant whose value is chosen to be 1. Moreover
\begin{equation}\label{23}
Q=-g^{\alpha\beta}(\mathcal{L}^{\rho}_{\mu\alpha}\mathcal{L}^{\mu}_{\beta\rho}
-\mathcal{L}^{\rho}_{\mu\rho}\mathcal{L}^{\mu}_{\alpha\beta}),
\end{equation}
where
\begin{equation}\label{24}
\mathcal{L}^{\rho}_{\mu\eta}=-\frac{1}{2}g^{\rho\lambda}
(\nabla_{\eta}g_{\mu\lambda}+\nabla_{\mu}g_{\lambda\eta}
-\nabla_{\lambda}g_{\mu\eta}),
\end{equation}
and the traces of the non-metricity tensor are defined by
\begin{align}\label{25}
Q_{\rho}= Q^{~\alpha}_{\rho~\alpha}, \quad \tilde{Q}_{\rho}=
Q^{\alpha}_{\rho\alpha}.
\end{align}
The superpotential in view of $Q$ is written as
\begin{align}\label{26}
P^{\rho}_{\alpha\beta}&=-\frac{1}{2}\mathcal{L}^{\rho}_{\alpha\beta}
+\frac{1}{4}(Q^{\rho}-\tilde{Q}^{\rho}) g_{\alpha\beta}- \frac{1}{4}
\delta ^{\rho}_{~[\alpha Q_{\beta}]}.
\end{align}
Furthermore, the expression of $Q$ obtained using the superpotential
(details are given in Appendix \textbf{A}) becomes
\begin{align}\label{27}
Q=-Q_{\rho\alpha\beta}P^{\rho\alpha\beta}
=-\frac{1}{4}(-Q^{\rho\beta\zeta}Q_{\rho\beta\zeta}
+2Q^{\rho\beta\zeta}Q_{\zeta\rho\beta}
-2Q^{\zeta}\tilde{Q}_{\zeta}+Q^{\zeta}Q_{\zeta}).
\end{align}

Taking the variation of $S$ with respect to the metric tensor as
zero yields the field equations
\begin{align}\nonumber
\delta S&=0=\int \frac{1}{2}\delta
[f(Q,T)\sqrt{-g}]+\delta[L_{m}\sqrt{-g}]d^{4}x \\\nonumber 0&=\int
\frac{1}{2}\bigg( \frac{-1}{2} f g_{\alpha\beta} \sqrt{-g} \delta
g^{\alpha\beta} + f_{Q} \sqrt{-g} \delta Q + f_{T} \sqrt{-g} \delta
T\bigg)\\\label{28}&-\frac{1}{2} T_{\alpha\beta} \sqrt{-g} \delta
g^{\alpha\beta}d^ {4}x.
\end{align}
The explicit formulation of $\delta Q$ is given in Appendix
\textbf{B}. Furthermore, we define
\begin{align}\label{29}
T_{\alpha\beta} &= \frac{-2}{\sqrt{-g}} \frac{\delta (\sqrt{-g}
L_{m})}{\delta g^{\alpha\beta}}, \quad \Theta_{\alpha\beta}=
g^{\rho\mu} \frac{\delta T_{\rho\mu}}{\delta g^{\alpha\beta}},
\end{align}
which implies that $ \delta T=\delta
(T_{\alpha\beta}g^{\alpha\beta})=(T_{\alpha\beta}+
\Theta_{\alpha\beta})\delta g^{\alpha\beta}$. Inserting the
aforementioned factors in Eq.\eqref{28}, we have
\begin{eqnarray}\nonumber
\delta S &=&0=\int \frac{1}{2}\bigg\{\frac{-1}{2}f
g_{\alpha\beta}\sqrt{-g} \delta g^{\alpha\beta} +
f_{T}(T_{\alpha\beta}+ \Theta_{\alpha\beta})\sqrt{-g} \delta
g^{\alpha\beta}-f_{Q} \sqrt{-g}
\\\nonumber
&\times& (P_{\alpha\rho\mu} Q_{\beta}~^{\rho\mu}-
2Q^{\rho\beta}~_{\alpha} P_{\rho\mu\beta}) \delta
g^{\alpha\beta}+2f_{Q} \sqrt{-g} P_{\rho\alpha\beta} \nabla^{\rho}
\delta g^{\alpha\beta}+2f_{Q}\sqrt{-g}
\\\label{30}
&\times& P_{\rho\alpha\beta} \nabla^{\rho} \delta g^{\alpha\beta}
\bigg\}- \frac{1}{2} T_{\alpha\beta}\sqrt{-g} \delta
g^{\alpha\beta}d^ {4}x.
\end{eqnarray}
Integration of the term $ 2 f_{Q} \sqrt{-g}
P_{\rho\alpha\beta}\nabla^{\rho}\delta g^{\alpha\beta}$ along with
the boundary conditions takes the form $ -2 \nabla^{\rho} (f_{Q}
\sqrt{-g} P_{\rho\alpha\beta})\delta g^{\alpha\beta}$. The terms
$f_{Q}$ and $f_{T}$ represent partial derivatives with respect to
$Q$ and $T$, respectively. Finally, we obtain the field equations as
\begin{align}\nonumber
T_{\alpha\beta}&=\frac{-2}{\sqrt{-g}} \nabla_{\rho} (f_{Q}\sqrt{-g}
P^{\rho}_{\alpha\beta})- \frac{1}{2} f g_{\alpha\beta} + f_{T}
(T_{\alpha\beta} + \Theta_{\alpha\beta}) -f_{Q} (P_{\alpha\rho\mu}
Q_{\beta}~^{\rho\mu} \\\label{31} &-2Q^{\rho\mu}~_{\alpha}
P_{\rho\mu\beta}).
\end{align}
The covariant differentiation of Eq.\eqref{31} yields the
non-conservation equation as
\begin{align}\nonumber
\nabla_{\alpha}T^{\alpha}_{~\beta}&=\frac{1}{f_{T}-1}\bigg[\nabla_{\alpha}
\bigg(\frac{1}{\sqrt{-g}}\nabla_{\rho}H^{~\rho\alpha}_{\beta}\bigg)-\nabla_{\alpha}
(f_{T} \Theta^{\alpha}_{~\beta}) -\frac{1}{\sqrt{-g}}\nabla_{\rho}
\nabla_{\alpha}H^{~\rho\alpha}_{\beta}\\\label{32}
&-2\nabla_{\alpha}A^{\alpha}_{~\beta}
+\frac{1}{2}f_{T}\partial_{\beta}T\bigg],
\end{align}
where hyper-momentum tensor density is defined as
\begin{equation}\label{81-a}
H^{~\rho\alpha}_{\beta}=\frac{\sqrt{-g}}{2}f_{T}\frac{\delta
T}{\delta
\hat{\Gamma}^{\beta}~_{\rho\alpha}}+\frac{\delta\sqrt{-g}L_{m}}
{\delta\hat{\Gamma}^{\beta}~_{\rho\alpha}}.
\end{equation}

\section{Reconstruction of GDE $f(Q,T)$ Model}

We consider a spatially flat Friedmann-Robertson-Walker metric of
the form
\begin{equation}\label{33}
ds^{2}=-dt^{2}+ a^{2}(t)[dx^{2}+ dy^{2}+dz^{2}],
\end{equation}
with $a(t)$ is the scale factor. The isotropic matter configuration
containing four-velocity fluid $u_{\alpha}$, the usual matter
density ($\rho_{M}$) and pressure ($P_{M}$), respectively, are given
as
\begin{equation}\label{33a}
\tilde{T}_{\alpha\beta}=(\rho_{M}+P_{M})u_{\alpha}u_{\beta}+P_{M}g_{\alpha\beta},
\end{equation}
The modified Friedmann equations for $f(Q,T)$ gravity are
\begin{align}\label{34}
3H^2 &=\rho_{eff}=\rho_{M}+\rho_{GDE},
\\\label{34-a}
2\dot{H}+3H^2&=P_{eff}=P_{M}+P_{GDE},
\end{align}
where $\rho_{GDE}$ and $P_{GDE}$ are the DE density and pressure,
respectively, given as
\begin{align}\label{35}
\rho_{GDE}&=\frac{1}{2}f(Q,T)-Qf_{Q}-f_{T}(\rho_{M}+P_{M}),\\\label{36}
P_{GDE}&=-\frac{1}{2}f(Q,T)+2f_{QQ}H+2f_{Q}\dot{H}+Qf_{Q},
\end{align}
where $H =\frac{\dot{a}}{a}$ signifies the Hubble parameter and dot
demonstrates the derivative with respect to the cosmic time $t$. The
non-metricity $Q$ in terms of the Hubble parameter (details are
given in Appendix \textbf{C}) is given as
\begin{align}\label{86}
Q=6H^{2}.
\end{align}
For an ideal fluid configuration, the conservation equation
\eqref{32} yields
\begin{align}\label{37}
\dot{\rho}_{M}+3H(\rho_{M}+P_{M})&=\frac{1}{(f_{T}-1)}
\bigg[2\nabla_{\beta}\mu^{\beta}(P_{M}f_{T})+f_{T}\nabla_{\beta}\mu^{\beta}T+
2T_{\alpha\beta}\nabla^{\alpha}\mu^{\beta} f_{T}\bigg].
\end{align}
According to the field equation \eqref{34}, we have
\begin{equation}\label{38}
\Omega_{M}+\Omega_{GDE}=1,
\end{equation}
where $\Omega_{M}=\frac{\rho_{M}}{3H^2}$ and
$\Omega_{GDE}=\frac{\rho_{GDE}}{3H^2}$ are the fractional energy
densities associated with usual matter and dark source,
respectively. Dynamical DE models involving energy density
proportional to the Hubble parameter are essential to explain the
accelerated expansion of the universe. In this scenario, the GDE
model is a dynamical DE model possessing energy density as
\begin{equation}\label{39}
\rho_{GDE}=\alpha H,
\end{equation}
where $\alpha$ is an arbitrary constant. In the following sections,
we recreate the GDE $f(Q,T)$ model using a correspondence technique
in the context of dust fluid $(P=0)$.

\subsection{Non-interacting GDE $f(Q,T)$ Model}

For the sake of simplicity, the standard $f(Q,T)$ function of the
following form is considered \cite{75}
\begin{equation}\label{40}
f(Q,T)=f_{1}(Q)+f_{2}(T),
\end{equation}
where $f_{1}$ and $f_{2}$ depend upon $Q$ and $T$, respectively. One
can observe that curvature and matter constituents are minimally
coupled in this scenario. This version of generic function shows
that the interaction is entirely gravitational and easy to manage.
This may effectively elaborate the ongoing expansion of the
universe. Furthermore, the reconstruction methodology reveals that
such generated models are physically feasible \cite{20,29}. Using
dust fluid and Eq.\eqref{40}, the field equations \eqref{34} and
\eqref{34-a} yield
\begin{equation}\label{41}
3H^{2}=\rho_{eff}=\rho_{M}+\rho_{GDE}, \quad
2\dot{H}+3H^{2}=P_{eff}=P_{GDE},
\end{equation}
where
\begin{align}\label{42}
\rho_{GDE}&=\frac{1}{2}f_{1}(Q)+\frac{1}{2}f_{2}(T)-Q f_{1Q}+
f_{2T}\rho_{M}, \\\label{43}
P_{GDE}&=-\frac{1}{2}f_{1}(Q)-\frac{1}{2}f_{2}(T)+2f_{1Q}\dot{H}+2
f_{1QQ}H+Qf_{1Q}.
\end{align}
The associated conservation equation \eqref{37} reduces to
\begin{align}
\dot{\rho}_{M}+3H\rho_{M}=\frac{1}{f_{2T}-1}\big[2Tf_{2TT}+f_{2T}\dot{T}\big].
\end{align}
This equation shows consistency with the standard continuity
equation if the right hand side is taken to be zero, implying
\begin{equation}\label{45}
{\dot\rho_{M}}+3H\rho_{M}=0\quad\Longrightarrow\quad
\rho_{M}=\rho_{0}a(t)^{-3},
\end{equation}
with the constraint
\begin{equation}\label{46}
f_{2T}+2Tf_{2TT}= 0,
\end{equation}
whose solution provides
\begin{equation}\label{47}
f_{2}(T)=\gamma_{1}T^\frac{1}{2}+\gamma_{2},
\end{equation}
where $\gamma_{1}$ and $\gamma_{2}$ are integration constants. We
equate Eqs.\eqref{39} and \eqref{42} along with constraint on
$f_{2}(T)$ \eqref{47} to establish a reconstruction paradigm with
the help of correspondence approach. The resulting differential
equation in $f_{1}(Q)$ is written as
\begin{equation}\label{48}
\frac{f_{1}(Q)}{2}-Q
f_{1Q}+\gamma_{1}T^\frac{1}{2}+\frac{1}{2}\gamma_{2}=\alpha H.
\end{equation}

In order to solve this differential, we consider the power-law
solution for the scale factor as
\begin{equation}\label{49}
a(t)=a_{0}t^m,   \quad m > 0,
\end{equation}
where $a_{0}$ indicates the current value of the scale factor. Using
the above relation, the expressions for Hubble parameter, its
derivative and non-metricity scalar in terms of cosmic time $t$
become
\begin{equation*}
H=\frac{m}{t}, \quad\dot{H}=-\frac{m}{t^{2}}, \quad Q=6
\frac{m^2}{t^2}.
\end{equation*}
Substituting \eqref{49} in \eqref{45}, it follows that
\begin{equation}\label{a}
\rho_{M}=\rho_{0}(a_{0}t^m)^{-3}.
\end{equation}
We are now able to derive the function $f_{1}(Q)$ by inserting
Eq.\eqref{49} in \eqref{48}, which turns out to be
\begin{align}\nonumber
f_{1}(Q)&=\frac{1}{3(4+3m)}-2\bigg(\frac{Q}{m^{2}}\bigg)^{\frac{-3m}{8}}
\bigg[-2^{3+\frac{3m}{8}}3^{1 +\frac{3m}{8}}\gamma_{1}
\\\label{50}
&+(4+3m)\bigg\{6^{\frac{3}{4}}\alpha\sqrt{m}+3\gamma_{2}
Q^{\frac{3}{4}}\bigg\}Q^{\frac{1}{4}}\bigg(\frac{Q}{m^{2}}\bigg)^{\frac{3m}{8}}\bigg]+
\sqrt{Q}c_{1},
\end{align}
where $c_{1}$ is the integration constant. As a result, the
reconstructed $f(Q,T)$ model is achieved by substituting
Eqs.\eqref{47} and \eqref{50} in \eqref{40} as
\begin{eqnarray}\nonumber
f(Q,T)&=&\frac{1}{3(4+3m)}-2\bigg(\frac{Q}{m^{2}}\bigg)^{\frac{-3m}{8}}
\bigg[-2^{3+\frac{3m}{8}}3^{1
+\frac{3m}{8}}\gamma_{1}+(4+3m)\bigg\{6^{\frac{3}{4}}\alpha\\\label{51}
&\times&\sqrt{m}+3\gamma_{2}
Q^{\frac{3}{4}}\bigg\}Q^{\frac{1}{4}}\bigg(\frac{Q}{m^{2}}\bigg)^
{\frac{3m}{8}}\bigg]+
\sqrt{Q}c_{1}+\gamma_{1}\sqrt{\frac{\rho_{0}}{a_{0}^{3}}}+\gamma_{2}.
\end{eqnarray}
The deceleration parameter is expressed as
\begin{equation}\label{67}
q=-\frac{a\ddot{a}}{\dot{a}^{2}}=-1+\frac{1}{m}.
\end{equation}
In relation with the deceleration parameter, the cosmic scale factor
is delineated as
\begin{equation}\label{68}
a(t)=t^{\frac{1}{(1+q)}},
\end{equation}
where $a_0$ is taken to be unity. Notice that the power-law model
corresponds to the expanding universe when $q>-1$.

The expanding phase as well as the current cosmic evolution,
respectively, are defined as
\begin{equation}\label{69}
H=\bigg(\frac{1}{1+q}\bigg)t^{-1}, \quad
H_{0}=\bigg(\frac{1}{1+q}\bigg)t^{-1}_{0}.
\end{equation}
According to the power-law cosmology, the expansion history of the
universe depends on two basic parameters, namely $H_{0}$ and $q$. By
considering the relationship between the scale factor $a$ and
redshift $z$, we have
\begin{equation}\label{70}
H=H_{0}\Psi^{1+q},
\end{equation}
where $\Psi=1+z$. Using Eq.\eqref{70} in \eqref{86}, $Q$ appears as
\begin{align}\label{71}
Q=6H_{0}^{2}\Psi^{2+2q}.
\end{align}
The reconstructed GDE $f(Q,T)$ model against red-shift parameter is
obtained by substituting the above relation in \eqref{51}, which
leads to
\begin{eqnarray}\nonumber
f(Q,T)&=&\gamma_{2}+\gamma_{1}\sqrt{\frac{\rho_{0}}{a_{0}^{3}}}+\sqrt{6}c_{1}\sqrt{\Psi^{2+2q}
H_{0}^{2}}-\bigg[2\bigg(\frac{\Psi^{2+2q}H_{0}^{2}}{m^{2}}\bigg)^{-3\frac{m}{8}}
\\\nonumber
&\times&\bigg\{-24\gamma_{1}+6(4+3m)
\big(\Psi^{2+2q}H_{0}^{2}\big)^{\frac{1}{4}}\bigg(\frac{\Psi^{2+2q}
H_{0}^{2}}{m^{2}}\bigg)^{3\frac{m}{8}}\\
\label{52}&\times&\big(\sqrt{m}\alpha
+3\gamma_{2}(\Psi^{2+2q}H_{0}^{2})^{\frac{3}{4}}\big)\bigg\}\bigg]\bigg
[3(4+3m)\bigg]^{-1}.
\end{eqnarray}
For the graphical analysis, we have specified the free parameters
$\gamma_{1}= -0.40,\\ \gamma_{2} =-0.001, c_{1} =2,$ and
$\rho_{0}=-1$ by assuming three specific values of $m = 1.2, 1.3$,
and $1.4$. The graphical behavior of the reconstructed $f(Q,T)$
model against red-shift factor is shown in Figure \textbf{1}. It is
seen that the reconstructed $f(Q,T)$ model grows gradually with the
increasing red-shift parameter. Moreover, an interesting result is
followed from $_{z\rightarrow 0}^{\lim}f(Q,T)=0$ which provides the
confirmation of the realistic model.
\begin{figure}\center
\epsfig{file=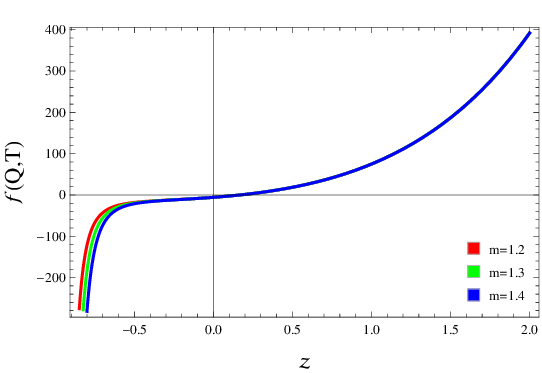,width=.5\linewidth} \caption{Plot of GDE
$f(Q,T)$  model against $z$.}
\end{figure}

\section{Cosmological Analysis}

In this section, we will discuss how the universe has evolved
through various phases. For this purpose, we employ the
reconstructed GDE $f(Q,T)$ model in the non-interacting case
\eqref{52}. We also illustrate the behavior of some cosmological
parameters such as the EoS parameter, state finder and
$\omega_{GDE}-\omega'_{GDE}$ planes. The stability of this model is
also discussed.

\subsection{Equation of State Parameter}

The EoS of DE can describe the cosmic inflation and expansion of the
universe. The condition for an accelerating cosmos is analyzed for
$\omega<-\frac{1}{3}$. When $\omega=-1$, it simply corresponds to
the cosmological constant, however, the values of
$\omega=\frac{1}{3}$ and $\omega=0$ demonstrate the radiation and
matter-dominated universe, respectively. Furthermore, phantom
scenario is observed by assuming $\omega<-1$. The EoS parameter is
expressed by
\begin{equation}\label{53}
\omega_{GDE}=
\frac{P_{eff}}{\rho_{eff}}=\frac{P_{GDE}}{\rho_{GDE}+\rho_{M}}.
\end{equation}
Equations \eqref{42}, \eqref{43} and \eqref{52} are utilized in the
above expression to evaluate the corresponding parameter associated
with the reconstructed model as
\begin{align}\nonumber
\omega_{GDE}&=\bigg[\Psi^{-2(1+q)}\bigg\{-m^{\frac{5}{2}}
\alpha\big(\Psi^{2+2q}H_{0}^{2}\big)^{\frac{1}{4}}
-9m^{3}\bigg(\frac{\Psi^{2+2q}\beta^2}{m^{2}}\bigg)^{\frac{-3m}{8}}
+72\sqrt{m}\alpha\big(\Psi^{2+2q}
\\\nonumber&\times H_{0}^{2}\big)^{\frac{5}{4}}
+m^{\frac{3}{2}}\alpha\big(\Psi^{2+2q} H_{0}^{2}\big)^{\frac{1}{4}}
(54\Psi^{2+2q}H_{0}^{2})
+12\Psi^{2+2q}H_{0}^{2}\bigg\{-1-\sqrt{\frac{\rho_{o}}{a_{o}^{3}}}
\Psi^{2+2q}
\\\nonumber
&\times H_{0}^{2}-4\bigg(\frac{(\Psi)^{2 + 2
q}H_{0}^{2}}{m^{2}}\bigg)^{\frac{-3m}{8}}
-2(6\Psi^{2+2q}H_{0}^{2})^{\frac{1}{2}}\}+12 m^{2}\bigg\{
\bigg(\frac{(\Psi)^{2+2q}H_{0}^{2}}{m^{2}}\bigg)^{\frac{-3m}{8}}
\\\nonumber
&\times6(\Psi)^{2+2q}H_{0}^{2}\bigg\}+m\Psi^{2+2q}H_{0}^{2}\bigg\{87-
9\sqrt{\frac{\rho_{o}}{a_{o}^{3}}}+324\Psi^{2+2q}H_{0}^{2}
(6\Psi^{2+2q}H_{0}^{2})^{\frac{1}{2}}
\\\nonumber
&+36\bigg(\frac{\Psi^{2+2q}H_{0}^{2}}{m^{2}}\bigg)^{\frac{-3
m}{8}}\bigg\}\bigg\}\bigg]\bigg[6(4
+3m)H_{0}^{2}\bigg\{\frac{\rho_{o}}{a_{o}^{3}}
+\frac{1}{2}\bigg\{\Psi^{2+2q}H_{0}^{2}+1-2\sqrt{m}\alpha\\
\label{54} &\times\bigg(\Psi^{2+2q}H_{0}^{2}\bigg)^{\frac{1}{4}}
-4\bigg(\frac{\Psi^{2 +
2q}H_{0}^{2}}{m^{2}}\bigg)^{\frac{-3m}{8}}\bigg\}\bigg\}\bigg]^{-1}.
\end{align}
Figure \textbf{2} illustrates the behavior of EoS parameter against
$z$ from which one can find that phantom epoch for current as well
as late time cosmic evolution is seen.
\begin{figure}\center
\epsfig{file=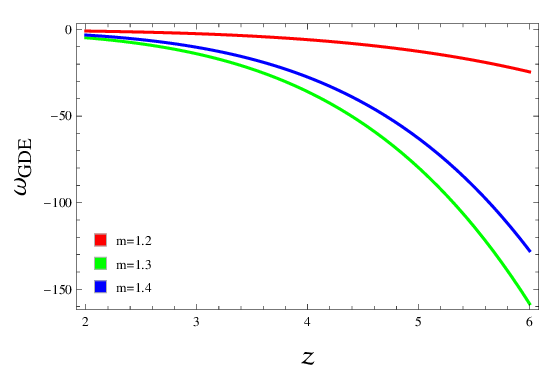,width=.5\linewidth}\caption{Plot of EoS
parameter against $z$.}
\end{figure}

\subsection{$\omega_{GDE}-\omega'_{GDE}$ Plane}

Here, we utilize the $\omega_{GDE}-\omega'_{GDE}$ phase plane, where
$\omega'_{GDE}$ indicates the evolutionary mode of $\omega_{GDE}$
and prime denotes the derivative with respect to $Q$. This cosmic
plane was established by Caldwell and Linder \cite{32} to
investigate the quintessence DE paradigm which can be split into
freezing $(\omega_{GDE} <0,~\omega'_{GDE}<0 )$ and thawing
$(\omega_{GDE}<0,~\omega'_{GDE}>0 )$ regions. In order to illustrate
the current cosmic expansion paradigm, the freezing region indicates
more accelerated phase as compared to thawing. The cosmic
trajectories of $\omega_{GDE}-\omega'_{GDE}$ plane for specific
choices of $m$ are shown in Figure \textbf{3} which provides the
freezing region of the cosmos. The expression of $\omega'_{GDE}$ is
given in Appendix \textbf{D}.
\begin{figure}\center
\epsfig{file=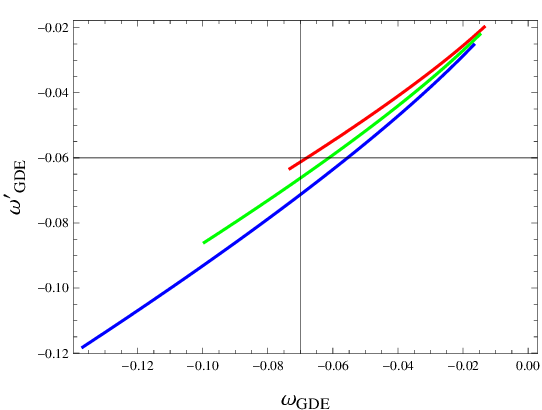,width=.56\linewidth}\caption{Plot of
$\omega_{GDE}-\omega'_{GDE}$ versus $z$.}
\end{figure}

\subsection{State Finder Analysis}

One of the techniques to examine the dynamics of the cosmos using a
cosmological perspective is state finder analysis. It is an
essential approach for understanding numerous DE models. As a
combination of the Hubble parameter and its temporal derivatives,
Sahni et al. \cite{33} established two dimensionless parameters
$(r,s)$ given as
\begin{equation}\label{64}
r=\frac{\dddot{a}}{a H^{3}}, \quad s=\frac{r-1}{3(q-\frac{1}{2})}.
\end{equation}
The acceleration of cosmic expansion is specified by parameter $s$
whereas the deviation from pure power-law behavior is specifically
defined by the parameter $r$. It is a geometric diagnostic that does
not support any specific cosmological paradigm. Secondly, it is such
an approach that does not depend upon any specific model to
distinguish between numerous DE scenarios, i.e., CG (Chaplygin gas),
HDE (Holographic DE), SCDM (standard CDM) and quintessence.

Several DE scenarios for the appropriate choices of $r$ and $s$
parametric values are given below.
\begin{itemize}
\item When $r=1$, $s=0$, it indicates the CDM model.
\item If $r=1$, $s=1$, then it denotes SCDM epoch.
\item When $r=1$, $s=\frac{2}{3}$, this epoch demonstrates the HDE
model.
\item When we have $r>1$, $s<0$, we get CG scenario.
\item Lastly, $r<1$, $s>0$ corresponds to quintessence paradigm.
\end{itemize}
For our considered setup, the parameters $r$ and $s$ in terms of
required factors are given in Appendix \textbf{E}. The left plot of
Figure \textbf{4} indicates the behavior of $r$ versus $z$ while the
right plot analyzes $s$ and $z$. The graphical analysis of $r-s$
phase plane in Figure \textbf{5} gives $r>1$ and $s<0$, indicating
the CG model.
\begin{figure}\center
\epsfig{file=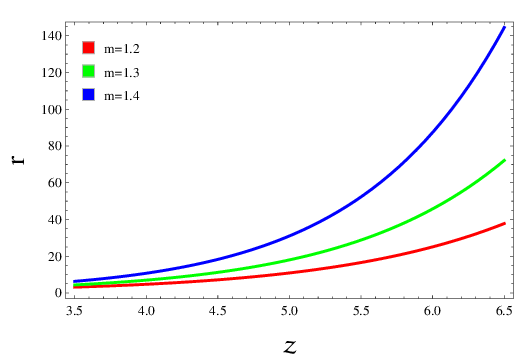,width=.5\linewidth}\epsfig{file=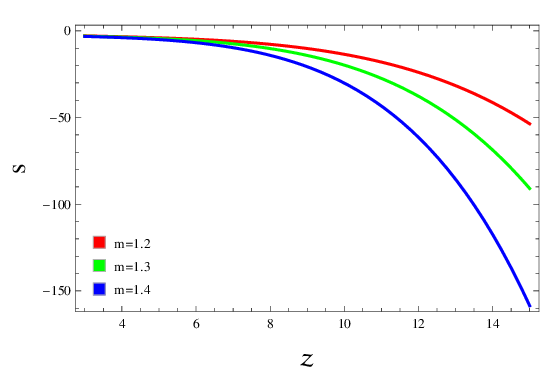,width=.5\linewidth}
\caption{Plots of $r$ against $z$ (left) and $s$ against $z$
(right).}
\end{figure}
\begin{figure}\center
\epsfig{file=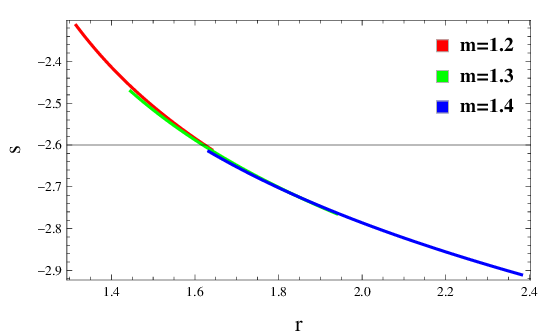,width=.5\linewidth}\caption{Plot of $r-s$ plane
against $z$.}
\end{figure}

\subsection{Squared Speed of Sound Parameter}

Stability analysis is a vital concept in understanding the structure
and behavior of the cosmos. In this regard, we consider the squared
speed of sound parameter to examine the stability of the GDE
$f(Q,T)$ model given as
\begin{equation}\label{55}
\nu_{s}^{2}=\frac{P_{GDE}}{\rho'_{GDE}}\omega'_{GDE} +\omega_{GDE}.
\end{equation}
Its positive value corresponds to the stable configuration whereas
the negative value implies unstable behavior for the analogous
model. Substituting the required expressions on the right hand side
of the above equation for the reconstructed model, we obtain the
squared speed of sound parameter given in Appendix \textbf{E}.
Figure \textbf{6} shows that the speed component is positive for all
the assumed values of $m$ and thus the reconstructed GDE $f(Q,T)$
model is stable throughout the cosmic evolution.
\begin{figure}\center
\epsfig{file=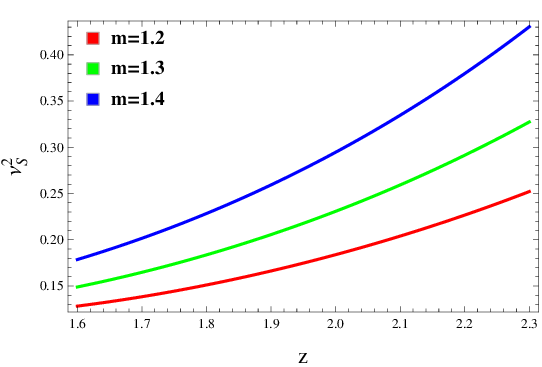,width=.5\linewidth}\caption{Plot of
$\nu_{s}^{2}$ against $z$ for non-interacting case.}
\end{figure}

\section{Final Remarks}

In the current scenario, DE is referred to as a vital representative
which tends to accelerate the expansion of the universe. In this
work, we have examined several characteristics of GDE model within
the context of recently developed $f(Q,T)$ theory of gravity. To
understand the role of $Q$ and $T$ in GDE ansatz, we have assumed a
particular $f(Q,T)=f_{1}(Q)+f_{2}(T)$ model. We have established a
reconstruction paradigm using a power-law scale factor as well as a
correspondence scenario for a flat FRW universe. We have then
investigated the EoS parameter, state finder,
$\omega_{GDE}-\omega'_{GDE}$ planes and squared sound speed analysis
for our derived model. For the graphical representation, we have
specified some values to the involved unknowns. The summary of the
results is given as follows.
\begin{itemize}
\item For non-interacting case, the reconstructed GDE $f(Q,T)$ model
reveals an increasing trend with convergence to zero from both sides
(Figure \textbf{1}).
\item The EoS parameter indicates an enormous phantom epoch of
the universe (Figure \textbf{2}).
\item The trajectories of the $\omega_{GDE}-\omega'_{GDE}$ plane
depicts the freezing region as both the parameters lie below the
zero range (Figure \textbf{3}).
\item The curvatures of $r-s$ plane illustrates
CG model because $r$ and $s$ have fulfilled the corresponding model
requirements (Figure \textbf{5}).
\item The squared speed of sound parameter indicates a stable regime
throughout the cosmic evolutionary paradigm (Figure \textbf{6}).
\end{itemize}

We have observed that for suitable choices of free parameters, our
reconstructed GDE $f(Q,T)$ model shows consistency with accelerated
expanding phenomenon of the universe. According to a review of
current observational data \cite{34}, the EoS parameter reveals a
stable trend when
\begin{align*}
\omega=-1.03\pm0.03 \quad (\text{Planck TT, TE, EE + LowE + lensing
+ SNe + BAO}).
\end{align*}
This value has been determined through several kinds of
observational techniques, with a 68$\%$ confidence level.
Furthermore, we have examined that the cosmic diagnostic state
finder parameters for our obtained model correspond to the
constraints and restrictions on the kinematics of the cosmos
\cite{35}. Sharif and Saba \cite {52} established the correspondence
of $f(G)$ theory with GDE paradigm and found that phantom phase for
the EoS parameter in non-interacting case. Our results are
consistent with these outcomes. Myrzakulov et al. \cite{51}
discussed the correspondence between $f(Q)$ and GDE model and found
that the reconstructed model is stable while the EoS parameters
crosses the phantom divide line. Our results are also consistent
with these consequences.

\section*{Appendix A: Evaluation of $Q=-Q_{\rho\alpha\beta}P^{\rho\alpha\beta}$}
\renewcommand{\theequation}{A\arabic{equation}}
\setcounter{equation}{0}

By considering Eq.\eqref{12}, we can write
\begin{align}\nonumber
-g^{\alpha\beta}\mathcal{L}^{\rho}_{\mu\alpha}\mathcal{L}^{\mu}_{\beta\rho}
&=-\frac{1}{4}g^{\alpha\beta}g^{\rho\lambda}g^{\mu\zeta}(Q_{\alpha\mu\lambda}
+Q_{\mu\lambda\alpha}-Q_{\lambda\alpha\mu})(Q_{\rho\beta\zeta}
+Q_{\beta\zeta\rho}-Q_{\zeta\rho\beta})
\\\nonumber
&=-\frac{1}{4}(Q^{\beta\zeta\rho}
+Q^{\zeta\rho\beta}-Q^{\rho\beta\zeta})(Q_{\rho\beta\zeta}
+Q_{\beta\zeta\rho}-Q_{\zeta\rho\beta})
\\\label{A1}
&=-\frac{1}{4}(2Q^{\rho\beta\zeta}Q_{\zeta\rho\beta}
-Q^{\rho\beta\zeta}Q_{\rho\beta\zeta}),
\\\nonumber
g^{\alpha\beta}\mathcal{L}^{\rho}_{\mu\rho}\mathcal{L}^{\mu}_{\alpha\beta}
&=\frac{1}{4}g^{\alpha\beta}g^{\mu\zeta}Q_{\mu}(Q_{\beta\alpha\zeta}
+Q_{\alpha\zeta\beta}-Q_{\zeta\beta\alpha})
=\frac{1}{4}Q^{\zeta}(2\tilde{Q}_{\zeta}-Q_{\zeta}),\\\label{A7}
\\
Q&=-\frac{1}{4}(-Q^{\rho\beta\zeta}Q_{\rho\beta\zeta}+2Q^{\rho\beta\zeta}Q_{\zeta\rho\beta}
-2Q^{\zeta}\tilde{Q}_{\zeta}+Q^{\zeta}{Q}_{\zeta}).\label{A8}
\end{align}
We can also find $Q$ as follows. Using Eq.\eqref{12} in \eqref{26},
we have
\begin{align}\label{A9}
P^{\rho\alpha\beta}&=&\frac{1}{4}\bigg[-Q^{\rho\alpha\beta}+Q^{\alpha\rho\beta}
+Q^{\beta\rho\alpha}+Q^{\rho}g^{\alpha\beta}-\tilde{Q}^{\rho}g^{\alpha\beta}
-\frac{1}{2}(g^{\rho\alpha}Q^{\beta}+g^{\rho\beta}Q^{\alpha}\bigg],
\end{align}
\begin{eqnarray}\nonumber
-Q_{\rho\alpha\beta}P^{\rho\alpha\beta}&=&-\frac{1}{4}
\bigg[-Q_{\rho\alpha\beta}Q^{\rho\alpha\beta}+Q_{\rho\alpha\beta}
Q^{\beta\rho\alpha}+Q_{\rho\alpha\beta}Q^{\beta\rho\alpha}
\\\nonumber
&+&Q_{\rho\alpha\beta}Q^{\rho}g^{\alpha\beta}-Q_{\rho\alpha\beta}
\tilde{Q}^{\rho}g^{\alpha\beta}-\frac{1}{2}Q_{\rho\alpha\beta}
(g^{\rho\alpha}Q^{\beta}+g^{\rho\beta}Q^{\alpha})\bigg]
\\\nonumber
&=&-\frac{1}{4}(-Q_{\rho\alpha\beta}Q^{\rho\alpha\beta}+2Q_{\rho\alpha\beta}
Q^{\alpha\rho\beta}+Q_{\rho}Q^{\rho}-2Q_{\rho}\tilde{Q^{\rho}})
\\\label{A10} &=&Q .
\end{eqnarray}
We have used the relations,
$Q_{\rho\alpha\beta}Q^{\alpha\rho\beta}=Q_{\rho\alpha\beta}Q^{\beta\rho\alpha}$
to obtain the above results since
$Q_{\rho\alpha\beta}Q^{\alpha\rho\beta}=Q_{\rho\beta\alpha}Q^{\alpha\rho\beta}
=Q_{\alpha\rho\beta}Q^{\rho\beta\alpha}=Q_{\rho\beta\alpha}Q^{\beta\alpha\rho}
Q_{\rho\alpha\beta}Q^{\beta\rho\alpha}$.

\section*{Appendix B: Steps for the Variation of $\delta Q$}
\renewcommand{\theequation}{B\arabic{equation}}
\setcounter{equation}{0}

The formulas related to non-metricity tensors used in the derivation
of $\delta Q$ are given as follows
\begin{align*}
Q_{\rho\alpha\beta}&=\nabla_{\rho}g_{\alpha\eta},
\\
Q^{\rho}_{~\alpha\beta}&=g^{\rho\mu}Q_{\mu\alpha\beta}=g^{\rho\mu}
\nabla_{\mu}g_{\alpha\beta}=\nabla^{\rho}g_{\alpha\beta},
\\
Q^{~~\mu}_{\rho~~\beta}&=g^{\alpha\zeta}Q_{\rho\zeta\beta}
=g^{\alpha\zeta}\nabla_{\rho}g_{\zeta\beta}
=-g_{\zeta\beta}\nabla_{\rho}g^{\alpha\zeta},
\\
Q_{\rho\alpha}^{~~\beta}&=g^{\beta\zeta}Q_{\rho\alpha\zeta}
=g^{\beta\zeta}\nabla_{\rho}g_{\alpha\zeta}
=-g_{\alpha\zeta}\nabla_{\rho}g^{\beta\zeta},
\\
Q^{\rho\alpha}_{~~\beta}&=g^{\rho\mu}g^{\alpha\zeta}\nabla_{\mu}g_{\zeta\beta}
=g^{\alpha\zeta}\nabla^{\rho}
g_{\zeta\beta}=-g_{\zeta\beta}\nabla^{\rho}g^{\alpha\zeta},
\\
Q^{\rho~~\beta}_{~\alpha}&=g^{\rho\mu}g^{\beta\zeta}\nabla_{\mu}g_{\alpha\zeta}
=g^{\beta\zeta}\nabla^{\rho}g_{\alpha\zeta}
=-g_{\alpha\zeta}\nabla^{\rho}g^{\beta\zeta},
\\
Q_{\rho}^{\quad\alpha\beta}&=g^{\alpha\zeta}g^{\beta\phi}\nabla_{\rho}g_{\zeta\phi}
=-g^{\mu\zeta}g_{\zeta\phi}\nabla_{\rho}g^{\beta\phi}
=-\nabla_{\rho}g^{\alpha\beta},
\\
Q^{\rho\alpha\beta}&=-\nabla^{\rho}g^{\alpha\beta}.
\end{align*}
Now, let us find the variation of $Q$ using Eq.\eqref{A8} as
\begin{align*}
\delta
Q&=-\frac{1}{4}\delta(-Q^{\rho\beta\zeta}Q_{\rho\beta\zeta}+2Q^{\rho\beta\zeta}
Q_{\zeta\rho\beta}-2Q^{\zeta}\tilde{Q}_{\zeta}+Q^{\zeta}Q_{\zeta})
\\
&=-\frac{1}{4}(-\delta
Q^{\rho\beta\zeta}Q_{\rho\beta\zeta}-Q^{\rho\beta\zeta}\delta
Q_{\rho\beta\zeta}+2\delta Q^{\rho\beta\zeta}Q_{\zeta\rho\beta}
\\
&+2Q^{\rho\beta\zeta}\delta Q_{\zeta\rho\beta}-2\delta
Q^{\zeta}\tilde{Q}_{\zeta}-2Q^{\zeta}\delta\tilde{Q}_{\zeta}+\delta
Q^{\zeta}Q_{\zeta}+Q^{\zeta}\delta Q_{\zeta}
\\
&=-\frac{1}{4}\bigg[Q_{\rho\beta\zeta}\nabla^{\rho}\delta
g^{\beta\zeta}-Q^{\rho\beta\zeta}\nabla_{\rho}\delta
g_{\beta\zeta}-2Q_{\zeta\rho\beta}\nabla^{\rho}\delta g^{\beta\zeta}
\\
&+2Q^{\rho\beta\zeta }\nabla_{\zeta}\delta
g_{\rho\beta}-2\hat{Q}_{\zeta}\delta(-g_{\alpha\beta}\nabla^{\zeta}g^{\alpha\beta})
-2Q^{\zeta} \delta(\nabla^{\lambda}g_{\zeta\lambda})
\\
&+Q_{\zeta}\delta(-g_{\alpha\beta}\nabla^{\zeta}g^{\alpha\beta})+Q^{\zeta}
\delta(-g_{\alpha\beta} \nabla_{\zeta }g^{\alpha\beta})\bigg]
\\
&=-\frac{1}{4}\bigg[Q_{\rho\beta\zeta}\nabla^{\rho}\delta
g^{\beta\zeta}-Q^{\rho\beta\zeta}\nabla_{\rho}\delta
g_{\beta\zeta}-2Q_{\zeta\rho\beta}\nabla^{\rho}\delta g^{\beta\zeta
}
\\
&+2Q^{\rho\beta\zeta}\nabla_{\zeta}\delta
g_{\rho\beta}+2\hat{Q}_{\zeta}\nabla^{\zeta }g^{\alpha\beta}\delta
g_{\alpha\beta}+2\hat{Q}_{\zeta}g_{\alpha\beta}\nabla^{\zeta}\delta
g^{\alpha\beta}
\\
&-2Q^{\zeta}\nabla^{\lambda}\delta
g_{\zeta\lambda}-Q_{\zeta}\nabla^{\zeta }g^{\alpha\beta}\delta
g_{\alpha\beta}-Q_{\zeta}g_{\alpha\beta}\nabla^{\zeta}\delta
g^{\alpha\beta}
\\
&-Q^{\zeta}\nabla_{\zeta}g^{\alpha\beta}\delta
g_{\alpha\beta}-Q^{\zeta} g_{\alpha\beta}\nabla_{\zeta}\delta
g^{\alpha\beta}\bigg].
\end{align*}
To simplify the above equation, we employ several useful formulas
given as
\begin{align*}
\delta g_{\alpha\beta}&=- g_{\alpha\rho}\delta
g^{\rho\mu}g_{\mu\beta},
\\
-Q^{\rho\beta\zeta}\nabla_{\rho}\delta g_{\beta\zeta}
&=-Q^{\rho\beta\zeta}\nabla_{\rho}(-g_{\beta\lambda}\delta
g^{\lambda\theta}g_{\theta\zeta})=2Q^{\rho\beta}_{\quad
\theta}Q_{\rho\beta\lambda}\delta
g^{\lambda\theta}+Q_{\rho\lambda\theta}\nabla^{\rho}g^{\lambda\theta}
\\
&=2Q^{\rho\phi}_{\quad\beta}Q_{\rho\phi\alpha}\delta
g^{\alpha\beta}+Q_{\rho\beta\zeta}\nabla^{\rho}g^{\beta\zeta},
\\
2Q^{\rho\beta\zeta}\nabla_{\zeta}\delta
g_{\rho\beta}&=-4Q_{\alpha}^{\quad\phi\zeta}Q_{\zeta
\phi\beta}\delta
g^{\alpha\beta}-2Q_{\beta\zeta\rho}\nabla^{\rho}\delta g^{\beta\zeta
},
\\
-2Q^{\zeta}\nabla^{\lambda}\delta
g_{\zeta\lambda}&=2Q^{\rho}Q_{\beta\rho\alpha}\delta
g^{\alpha\beta}+2Q_{\alpha}\tilde{Q}_{\beta}\delta g^{\alpha\beta}
+2Q_{\beta}g_{\rho\zeta}\nabla^{\rho}g^{\beta\zeta}.
\end{align*}
Thus, we have
\begin{align*}
\delta Q=2P_{\rho\beta\zeta
}\nabla^{\rho}g^{\beta\zeta}-(P_{\alpha\rho\mu}
Q_{\beta}^{\quad\alpha\beta}-2Q^{\alpha\beta}_{\quad\alpha}P_{\rho\mu\beta})\delta
g^{\alpha\beta},
\end{align*}
where
\begin{align}\label{B15}
2P_{\rho\beta\zeta}&=-\frac{1}{4}\bigg[2Q_{\rho\beta\zeta}-2Q_{\zeta\rho\beta}
-2Q_{\beta\zeta\rho}+2(\tilde Q_{\rho}-
Q_{\rho})g_{\beta\zeta}+2Q_{\beta}g_{\rho\zeta}\bigg],
\end{align}
and
\begin{align}\nonumber
4(P_{\alpha\rho\mu}Q_{\beta}^{\quad\alpha\beta}
-2Q^{\rho\mu}_{\quad\alpha}P_{\rho\mu\beta})&
=2Q^{\alpha\beta}_{\quad\beta}
Q_{\rho\mu\beta}-4Q_{\alpha}^{\quad\alpha\beta}Q_{\mu\rho\beta}
+2\tilde Q^{\rho}Q_{\rho\alpha\beta}\
\\\label{B16}
&+2Q^{\rho}Q^{\beta\rho\alpha}+2Q_{\alpha}\tilde
Q_{\beta}-Q^{\rho}Q_{\rho\alpha\beta}.
\end{align}

\section*{Appendix C: Calculation of $Q=6H^{2}$}
\renewcommand{\theequation}{C\arabic{equation}}
\setcounter{equation}{0}

Using Eq.\eqref{A10}, we have
\begin{equation}\label{E1}
Q=-\frac{1}{4}\big[-Q_{\rho\alpha\beta}Q^{\rho\alpha\beta}
+2Q_{\rho\alpha\beta}Q^{\alpha\rho\beta}
-2\tilde{Q}^{\rho}Q_{\rho}+Q^{\rho}Q_{\rho}\big],
\end{equation}
in which
\begin{align*}
-Q_{\rho\alpha\beta}Q^{\rho\alpha\beta}
&=\nabla_{\rho}g_{\alpha\beta}\nabla^{\rho}g^{\alpha\beta}=12H^{2},
\\
Q_{\rho\alpha\beta}Q^{\alpha\rho\beta}&=-\nabla_{\rho}g_{\alpha\beta}
\nabla^{\alpha}g^{\rho\beta}=0,
\\
\tilde{Q}^{\rho}Q_{\rho}&=(g_{\alpha\zeta}\nabla_{\rho}g^{\alpha\zeta})
(\nabla_{\mu}g^{\rho\mu})=0,
\\
Q^{\rho}Q_{\rho}&=(g_{\zeta\alpha}\nabla_{\rho}g^{\zeta\alpha})
(g_{\phi\beta}\nabla^{\rho}g^{\zeta\beta})=-36H^{2}.
\end{align*}
Substituting all the above expressions in \eqref{E1}, it follows
that
\begin{align*}
Q&= 6H^{2}.
\end{align*}

\section*{Appendix D: Calculation of $\omega'_{GDE}$}
\renewcommand{\theequation}{D\arabic{equation}}
\setcounter{equation}{0}

\begin{align}\nonumber
\omega'_{GDE}=&-\bigg[\Psi^{-2(1+q)}\bigg\{-\frac{3m^{\frac{5}{2}}
(2+2q)\Psi^{1+2q}
\alpha\beta^{2}}{4\big(\Psi^{2+2q}H_{0}^{2}\big)^{\frac{3}{4}}}+90\sqrt{m}
(2+2q)\Psi^{1+2q} \alpha H_{0}^{2}
\\\nonumber
&\times\big(\Psi^{2+2q}H_{0}^{2}\big)^{\frac{1}{4}}+54m^{\frac{3}{2}}(2+2q)
\Psi^{1+2q}\alpha H_{0}^{2}(\Psi^{2+2q}H_{0}^{2})^{\frac{1}{4}}
+m^{2}\Psi^{1+2q}H_{0}^{2}
\\\nonumber
&\times(2+2q)\big(\Psi^{2+2q}H_{0}^{2}{m^{2}} \big)^{\frac{-3m}{8}}
+m^{\frac{3}{2}}(2+2q) \Psi^{1+2q}\alpha H_{0}^{2}
(-4+54\Psi^{2+2q}H_{0}^{2})
\\\nonumber &
\times\bigg\{\big(\Psi^{2+2q}4H_{0}^{2}\big)^{\frac{3}{4}}
+m\Psi^{2+2q}H_{0}^{2}
\bigg\{\sqrt{6}(2+2q)\Psi^{1+2q}H_{0}^{2}\sqrt{\Psi^{2+2q}
H_{0}^{2}}\Psi^{1+2q}
\\\nonumber
&\times H_{0}^{2}-\frac{1}{2m} 27(2+2q)
\bigg(\frac{\Psi^{2+2q}H_{0}^{2}}{{m^{2}}}\bigg)^{-1-\frac{3m}{8}}\bigg\}\Psi^{1+2q}
H_{0}^{2} +12m^{2}\bigg\{6(2+2q)
\\\nonumber
&\times\Psi^{1+2q}H_{0}^{2}+\frac{1}{8m}3(2+2q)H_{0}^{2}\Psi^{1+2q}
\Psi^{1+2q}\bigg(\frac{\Psi^{2+2q}H_{0}^{2}}{{m^{2}}}
\bigg)^{{-1-\frac{3m}{8}}}\bigg\} +12\Psi^{2+2q}
\\\nonumber
&\times H_{0}^{2}\bigg\{(2 + 2
q)\Psi^{1+2q}H_{0}^{2}-\frac{\sqrt{6}(2+2q)\Psi^{1+ 2q}H_{0}^{2}}
{\sqrt\Psi^{2+2q}H_{0}^{2}} +\frac{3}{2m}(2+2q)\Psi^{1+2q}
\\\nonumber
&\times H_{0}^{2}
\bigg(\frac{\Psi^{2+2q}H_{0}^{2}}{m^{2}}\bigg)^{-1-\frac{3m}{8}}\bigg\}
+\Psi^{2+2q}H_{0}^{2}\bigg\{36(2+2q)\Psi^{1+2q}
+\bigg(\frac{\Psi^{2+2q}H_{0}^{2}}{{m^{2}}} \bigg)^{\frac{-3m}{8}}
\\\nonumber
&\times3\Psi^{1+2q} H_{0}^{2}-
\frac{\sqrt{6}(2+2q)\Psi^{1+2q}H_{0}^{2}}{\sqrt{\Psi^{2+2q}
H_{0}^{2}}}(2+2q)\frac{1}{2m}H_{0}^{2}\bigg\} +12(2+2q)\Psi^{1+2q}
\\\nonumber
&\times H_{0}^{2}\bigg\{1-\sqrt{\frac{\rho_{o}}{a_{o}^{3}}}
+\Psi^{2+2q} +36H_{0}^{2}-2\sqrt{6}\sqrt{\Psi^{2+2q}H_{0}^{2}}
-4\bigg(\frac{\Psi^{2+2q}H_{0}^{2}}{m^{2}}\bigg)^{\frac{-3
m}{8}}\bigg\}
\\\nonumber
&+m(2+2q)
\Psi^{1+2q}H_{0}^{2}\bigg\{87-9\sqrt{\frac{\rho_{o}}{a_{o}^{3}}}
+(\Psi)^{2+2q}H_{0}^{2}-18\sqrt{6}\sqrt{(\Psi)}^{2+2q}H_{0}^{2}
\\\nonumber
&+36\bigg(\frac{\Psi^{2+2q}H_{0}^{2}}{m^{2}}\bigg)^{-3\frac{m}{8}}\bigg\}\bigg\}\bigg]
\bigg[6(4+3m)H_{0}^{2}\bigg\{\frac{\rho_{o}}{a_{o}^{3}}+\bigg\{\frac{1}{2}
\bigg(-1-12H_{0}^{2}\Psi^{2+2q}
\\\nonumber
&+2\alpha\big(\Psi^{2+2q}H_{0}^{2}\big)^{\frac{1}{4}}
-4\bigg(\frac{\Psi^{2+2q}H_{0}^{2}}{m^{2}}\bigg)^{-3\frac{m}{8}}
\bigg\}\bigg\}\bigg\}\bigg]^{-1}
+\bigg\{\Psi^{-2(1+q)}\bigg\{\Psi^{1+2q}H_{0}^{2})
\\\nonumber
&\times(2+2q)+\frac{\sqrt{m}(2+2q) \Psi^{1+2q}\alpha
H_{0}^{2}}{2(\Psi^{2+2q}H_{0}^{2})^{\frac{3}{4}}}\bigg\}
\bigg\{\frac{1}{2m}(3(2+2q)\Psi^{1+2q} H_{0}^{2}-3m^{\frac{5}{2}}
\\\nonumber
&+\bigg(\frac{\Psi^{2+2q} H_{0}^{2}}{m^{2}}\bigg)^{-1\frac{-3m}{8}}
(\alpha\Psi^{2+2q}H_{0}^{2})^{\frac{1}{4}}
+\sqrt{m}\alpha(\Psi^{2+2q}H_{0}^{2})^{\frac{5}{4}}-m^{3}
\bigg(\frac{\Psi^{2+2q}H_{0}^{2}}{m^{2}}\bigg)^{\frac{-3m}{8}}
\\\nonumber
&+390m^{\frac{3}{2}}\alpha(\Psi^{2+2q}H_{0}^{2})^\frac{1}{4}
(-4+54\Psi^{2+2q}H_{0}^{2})+12\Psi^{2+2q}H_{0}^{2}\bigg\{-1
-\sqrt{\frac{\rho_{0}}{a_{0}^{3}}}
\\\nonumber
&+36\Psi^{2+2q}H_{0}^{2}-2\sqrt{6}\sqrt{\Psi^{2+2q}H_{0}^{2}}+72\alpha\sqrt{m}
-4\bigg(\frac{\Psi^{2+2q}H_{0}^{2}}{m^{2}}\bigg)^{-3\frac{m}{8}}\bigg\}
+\bigg\{\Psi^{2+2q}
\\\nonumber
&\times H_{0}^{2})-\bigg(\frac{\Psi^{2+2q}H_{0}^{2}}{m^{2}}
\bigg)^{\frac{-3m}{8}}\bigg\}
12m^{2}+m\Psi^{2+2q}H_{0}^{2}\bigg\{87-9\sqrt{\frac{\rho_{0}}{a_{0}^{3}}}
+324\Psi^{2+2q}
\\\nonumber
&\times H_{0}^{2}-18\sqrt{6}\sqrt{\Psi^{2+2q}H_{0}^{2}}
+36\bigg(\frac{\Psi^{2+2q}H_{0}^{2}}{m^{2}}\bigg)^{-3\frac{m}{8}}
\bigg\}\bigg\}\bigg\}+((1 + z)^{2+2q}H_{0}^{2})^{\frac{1}{4}}
\\\nonumber
&\times\bigg\{12(4+3m)H_{0}^{2}\bigg[\frac{\rho_{0}}{a_{0}^{3}}
+\frac{1}{2}\bigg\{-1-H_{0}^{2}+\alpha(\Psi^{2+2q}H_{0}^{2})^{\frac{1}{4}}
-4\bigg(\frac{\Psi^{2+2q}H_{0}^{2}}{m^{2}}\bigg)^{\frac{-3m}{8}}
\\\nonumber
&\times\Psi^{2+2q}\bigg\}
\bigg]^{2}\bigg\}+\bigg\{(1+q)\Psi^{-1-2(1+q)}\bigg\{-3m^{\frac{5}{2}}
\alpha\big(\Psi^{2+2q}H_{0}^{2}\big)^{\frac{1}{4}} +72\sqrt{m}\alpha
\\\nonumber
&\times\bigg(\Psi^{2+2q}H_{0}^{2}\bigg)^{\frac{5}{4}}
-\bigg(\frac{(\Psi)^{2+2q}H_{0}^{2}}{m^{2}}\bigg)^{\frac{-3m}{8}}
\times m^{\frac{3}{2}} \alpha ((1 +
z)^{2+2q}H_{0}^{2})^{\frac{1}{4}}(-4
\\\nonumber
&+54(\Psi)^{2+2q}H_{0}^{2}+ 12\sqrt{m}\alpha(\Psi)^{2+2q}H_{0}^{2}
\bigg\{-1
-\sqrt{\frac{\rho_{0}}{a_{0}^{3}}}+36\Psi^{2+2q}H_{0}^{2}-2
\\\nonumber
&\times\sqrt{6}\sqrt{(\Psi)^{2+2q}H_{0}^{2}}+12\sqrt{m}\alpha
-4\bigg(\frac{\Psi^{2+2q}H_{0}^{2}}{m^{2}}\bigg)^{-3\frac{m}{8}}\bigg\}+
\bigg\{6\Psi^{2+2q}H_{0}^{2}
\\\nonumber&-
\bigg(\frac{\Psi^{2+2q}H_{0}^{2}}{m^{2}}\bigg)^{-3\frac{m}{8}}
12m^{2}\bigg\}+m(\Psi)^{2+2q}H_{0}^{2}\bigg\{87
-9\sqrt{\frac{\rho_{0}}{a_{0}^{3}}} +324\sqrt{m}\Psi^{2+2q}
\\\nonumber &-18\sqrt{6}\sqrt{(\Psi)^{2+2q}\beta}^{2}H_{0}^{2}
+\times
36(\frac{((\Psi)^{2+2q}H_{0}^{2}}{m^{2}})^{-3\frac{m}{8}}\bigg\}\bigg\}\bigg\}
\bigg[3(4+3m)H_{0}^{2}
\\\nonumber
&\times \bigg\{\frac{\rho_{0}}{a_{0}^3}+\frac{1}{2}\bigg\{-1-
12(\Psi)^{2+2q}H_{0}^{2}
-4\bigg(\frac{\Psi^{2+2q}H_{0}^{2}}{m^{2}}\bigg)^{\frac{-3m}{8}}
-2\sqrt{m}\alpha((\Psi)^{2+2q}
\\\label{57}
&\times H_{0}^{2})^{\frac{1}{4}}\bigg\}\bigg\}\bigg]^{-1}.
\end{align}

\section*{Appendix E: Evaluation of $r$ and $s$ Parameters}
\renewcommand{\theequation}{E\arabic{equation}}
\setcounter{equation}{0}

\begin{align}\nonumber
r&=\frac{1}{4}\bigg[2-\bigg\{\Psi^{-2(1+q)}\bigg\{-3m^{\frac{5}{2}}\alpha(\Psi^{2+2q}
H_{0}^{2})^{\frac{1}{4}}
-9m^{3}\bigg(\frac{\Psi^{2+2q}H_{0}^{2}}{m^{2}}\bigg)^{-3\frac{m}{8}}
+(\Psi^{2+2q}
\\\nonumber
&\times H_{0}^{2})^{\frac{5}{4}}+72\sqrt{m}
\alpha+\big(-4+54\Psi^{2+2q}H_{0}^{2}\big)
m^{\frac{3}{2}}72\sqrt{\alpha}
\big((\Psi)^{2+2q}H_{0}^{2}\big)^{\frac{1}{4}}+12\Psi^{2+2q}
\\\nonumber
&\times H_{0}^{2}\bigg\{-1 -\sqrt{\frac{\rho_{o}}{a_{o}^{3}}}+
36m^2\Psi^{2+2q}H_{0}^{2}
-2(6(\Psi)^{2+2q}H_{0}^{2})^{\frac{1}{2}}-4\bigg(\frac{\Psi^{2+2q}H_{0}^{2}}
{m^{2}}\bigg)^{-3\frac{m}{8}}\bigg\}
\\\nonumber
&+12 m^{2}\bigg\{6(\Psi)^{2+2q}H_{0}^{2}-
\bigg(\frac{(\Psi)^{2+2q}H_{0}^{2}}{m^{2}}\bigg)^{-3\frac{m}{8}}\bigg\}+
m\Psi^{2+2 q}H_{0}^{2}\bigg\{87-\sqrt{(\Psi)^{2+2q}H_{0}^{2}}
\\\nonumber
&\times18\sqrt{6}-9\sqrt{\frac{\rho_{o}}{a_{o}^{3}}}+324\Psi^{2+2q}
H_{0}^{2}+36\bigg(\frac{(\Psi)^{2+2q}H_{0}^{2}}{m^{2}}
\bigg)^{-3\frac{m}{8}}\bigg\}\bigg\}\bigg\}
\bigg[(4+3m)H_{0}^{2}\bigg\{\frac{\rho_{o}}{a_{o}^{3}}
\\\nonumber
&+\frac{1}{2}\bigg\{-1-\Psi^{2+2q}H_{0}^{2}+\sqrt{m}\alpha((\Psi)^{2+2q}
H_{0}^{2})^{\frac{1}{4}}-\bigg((\Psi)^{2+2q}
H_{0}^{2}m^{-2}\bigg)^{-3\frac{m}{8}}\bigg\}\bigg\}\bigg]^{-1} +8
\\\nonumber &\times
\bigg\{\frac{1}{2}
-\bigg\{\Psi^{-2(1+q)}\bigg\{-3m^{\frac{5}{2}}\alpha
\big((\Psi)^{2+2q}H_{0}^{2}\big)^{\frac{1}{4}} +
\big(\Psi^{2+2q}H_{0}^{2}\big)^{\frac{5}{4}}
+\alpha\big(\Psi^{2+2q}H_{0}^{2}\big)^{\frac{1}{4}}
\\\nonumber &\times72\sqrt{m}\alpha
-m^{3}\big((\Psi)^{2+2q}H_{0}^{2}m^{-2}\big)^{\frac{-3m}{8}}(-4 +
54\Psi^{2+2q}H_{0}^{2})+12(\Psi)^{ 2+ 2 q}H_{0}^{2} \bigg\{-1
\\\nonumber
&-\sqrt{\frac{\rho_{o}}{a_{o}^{3}}}+36(\Psi)^{2+2q}H_{0}^{2}
-(6\Psi^{2+2q}H_{0}^{2})^{\frac{1}{2}} m^{\frac{3}{2}}
-\big(\Psi^{2+2q}\big)^{-3\frac{m}{8}}
\big((H_{0}^{2}m^{-2})^{-3\frac{m}{8}}\big)\bigg\}
\\\nonumber
&+ \bigg\{6\Psi^{2+2q}
H_{0}^{2}-\bigg(\frac{(\Psi)^{2+2q}H_{0}^{2}}{m^{2}}\bigg)^{-3\frac{m}{8}}\bigg\}12m^{2}
+m\Psi^{2+2q}H_{0}^{2}\bigg\{324\Psi^{2+2q}H_{0}^{2} -18
\\\nonumber
&\times(6(\Psi)^{2 + 2
q}H_{0}^{2})^{\frac{1}{2}}87-9\sqrt{\frac{\rho_{0}}{a_{0}^{3}}}
+36\bigg(\frac{(1 +
z)^{2+2q}H_{0}^{2}}{m^{2}}\bigg)^{-3\frac{m}{8}}\bigg\}\bigg\}\bigg\}
\times\bigg\{(4+3m)
\\\nonumber
&\times\bigg\{\frac{\rho_{0}}{a_{0}^{3}} +\frac{1}{2}\bigg\{-1-12
\Psi^{2+2q}H_{0}^{2}+ ((\Psi)^{2 + 2 q}
H_{0}^{2})^{\frac{1}{4}}-\bigg(\frac{(\Psi)^{2+2q}H_{0}^{2}}
{m^{2}}\bigg)^{-3\frac{m}{8}}\Psi^{2+2q}
\\\nonumber
&\times H_{0}^{2}\bigg\}\bigg\}\bigg\}^{-1}\bigg]^{2} -\bigg\{(1 +
q)(1 +
z)^{-4-3q}\bigg\{2\bigg\{-24(\Psi)^{2+2q}H_{0}^{2}m^{\frac{3}{2}}
\alpha+ \sqrt{m}\alpha
\\\nonumber
&\times\big(\Psi^{2 + 2 q}H_{0}^{2}\big)^{\frac{1}{4}} +3m
\bigg(\frac{\Psi^{2 + 2
q}H_{0}^{2}}{m^{2}}\bigg)^{-3\frac{m}{8}}\bigg\}\bigg\}\bigg\{-3m^{\frac{5}{2}}\alpha
\big((\Psi)^{2 + 2 q} H_{0}^{2}\big)^{\frac{1}{4}}+72\sqrt{m}\alpha
\\\nonumber
&\times \big((\Psi)^{2 + 2 q} H_{0}^{2}\big)^{\frac{5}{4}}-m^3
\bigg(\frac{(\Psi)^{2 + 2 q}
H_{0}^{2}}{m^{2}}\bigg)^{-3\frac{m}{8}}+\big((\Psi)^{2 + 2
q}\big)^{\frac{1}{4}}(-4+54 (\Psi)^{2 + 2 q}H_{0}^{2})
\\\nonumber
&+12(\Psi)^{2 + 2 q}H_{0}^{2}\bigg \{-2\sqrt{6}\sqrt{(\Psi)^{2 + 2
q}H_{0}^{2}} -\bigg\{12m^{2} -\sqrt{\frac{\rho_{0}}{a_{0}^{3}}} +
36(\Psi)^{2 + 2 q} H_{0}^{2}-4
\\\nonumber
&\bigg(\frac{\Psi^{2 + 2
q}H_{0}^{2}}{m^{2}}\bigg)^{-3\frac{m}{8}}\bigg\}+ 12m^{2}\bigg\{6
\Psi^{2 + 2 q}H_{0}^{2} -\bigg(\frac{(\Psi)^{2 + 2
q}H_{0}^{2}}{m^{2}}\bigg)^{-3\frac{m}{8}}\bigg\}+m\Psi^{2 + 2
q}H_{0}^{2}
\\\nonumber
&\times\bigg\{87-9\sqrt{\frac{\rho_{0}}{a_{0}^{3}}}+324(\Psi)^{2 + 2
q}H_{0}^{2} -18(6(\Psi)^{2 + 2 q}H_{0}^{2})^{\frac{1}{2}}
+\big((\Psi^{2 + 2 q}
H_{0}^{2}m^{-2}\big)^{\frac{-3m}{8}}\bigg\}\bigg\}
\\\nonumber
&+8\bigg\{{\frac{\rho_{0}}{a_{0}^{3}}}+\bigg\{(\Psi)^{2 + 2
q}H_{0}^{2}+2\sqrt{m}\alpha\big((\Psi)^{2 + 2
q}H_{0}^{2}\big)^{\frac{1}{4}}) -4\bigg(\frac{(\Psi)^{2 + 2
q}H_{0}^{2}}{m^{2}}\bigg)^{-3\frac{m}{8}}\bigg\}\bigg\}
\\\nonumber
&\times\bigg\{-3 m^{\frac{5}{2}} \alpha \big((\Psi)^{2 + 2 q}
H_{0}^{2}\big)^{\frac{1}{4}}+\sqrt{m}\alpha \big((\Psi)^{2+2q}
H_{0}^{2}\big)^{\frac{5}{4}}-m^{3}\bigg(\frac{(\Psi)^{2 + 2
q}H_{0}^{2}}{m^{2}}\bigg)^{-3\frac{m}{8}}
\\\nonumber
&+\big(\Psi^{2 + 2 q} H_{0}^{2}\big)^{\frac{1}{4}}(-4 + 54 (\Psi)^{2
+ 2 q}H_{0}^{2})+12 (\Psi)^{2 + 2 q}
H_{0}^{2}\bigg\{-1-\sqrt{\frac{\rho_{0}}{a_{0}^{3}}}+ 36(\Psi)^{2 +
2 q}
\\\nonumber &\times H_{0}^{2}-2\sqrt{6}\sqrt{(\Psi)^{2 + 2
q}H_{0}^{2}} -4 \bigg(\frac{(\Psi)^{2 + 2
q}H_{0}^{2}}{m^{2}}\bigg)^{-3\frac{m}{8}}\bigg\}+ 12m^{2}\bigg\{
\bigg(\frac{(\Psi)^{2 + 2 q}H_{0}^{2}}{m^{2}}\bigg)^{-3\frac{m}{8}}
\\\nonumber
&\times6\Psi^{2 + 2 q}H_{0}^{2})\bigg\}
+m\Psi^{2+2q}H_{0}^{2}\bigg\{87
-9\sqrt{\frac{\rho_{0}}{a_{0}^{3}}}+324\Psi^{2+2q}H_{0}^{2}
\sqrt{(\Psi)^{2+2q}H_{0}^{2}}-18\sqrt{6}
\\\nonumber
&+ 36\big((\Psi)^{2+2 q}
H_{0}^{2}m^{-2}\big)^{-3\frac{m}{8}}\bigg\}\bigg\}
-\bigg\{{\frac{\rho_{0}}{a_{0}^{3}}}+\frac{1}{2}\bigg\{-12 (\Psi)^{2
+ 2 q}H_{0}^{2})+ \big((\Psi)^{2 + 2 q} H_{0}^{2}\big)^{\frac{1}{4}}
\\\nonumber
&\times2\sqrt{m}\alpha- 4\bigg(\frac{(\Psi)^{2 + 2
q}H_{0}^{2}}{m^{2}}\bigg)^{-3\frac{m}{8}}\bigg\}\bigg\}\bigg\{-6
m^{\frac{5}{2}} \alpha\big((\Psi)^{2 + 2 q}
H_{0}^{2}\big)^{\frac{1}{4}} +\big(\Psi^{2 + 2
q}H_{0}^{2}\big)^{\frac{5}{4}}
\\\nonumber
&+720\sqrt{m}\alpha+m^{\frac{3}{2}}\alpha\big((\Psi)^{2 + 2 q}
H_{0}^{2}\big)^{\frac{5}{4}}+ m^{4}\bigg(\frac{(\Psi)^{2 + 2
q}H_{0}^{2}}{m^{2}}\bigg)^{-3\frac{m}{8}}+\alpha\big((\Psi)^{2 + 2
q}H_{0}^{2}\big)^{\frac{1}{4}}
\\\nonumber
&+4m^{\frac{3}{2}}(27 (\Psi)^{2 + 2 q}
 H_{0}^{2})+ 96(\Psi)^{2 + 2 q} H_{0}^{2}\bigg\{
-\sqrt{\frac{\rho_{0}}{a_{0}^{3}}} +36(\Psi)^{2 + 2 q}H_{0}^{2}-
\Psi^{2 + 2 q}H_{0}^{2})^{\frac{1}{2}}
\\\nonumber &-4
\bigg(\frac{(\Psi)^{2 + 2
q}H_{0}^{2}}{m^{2}}\bigg)^{-3\frac{m}{8}}\bigg\} + 8m(\Psi)^{2 + 2
q}H_{0}^{2}\bigg\{-\sqrt{\frac{\rho_{o}}{a_{o}^{3}}}-(6\Psi^{2 + 2
q}H_{0}^{2})^{\frac{1}{2}}+(\Psi)^{2 + 2 q}
\\\nonumber &
\times H_{0}^{2}+\big(\Psi^{2 + 2 q}
H_{0}^{2}m^{-2}\big)^{-3\frac{m}{8}}\bigg\}+m(\Psi)^{2 + 2 q}
H_{0}^{2}\bigg\{72(\Psi)^{2 + 2 q} H_{0}^{2}-\sqrt{(\Psi)^{2 + 2 q}
H_{0}^{2}}
\\\nonumber &\times2\sqrt{6}-3
m\bigg(\frac{(\Psi)^{2 + 2 q}
H_{0}^{2}}{m^{2}}\bigg)^{-3\frac{m}{8}}\bigg\}+m^{2}\bigg\{16
\Psi^{2 + 2 q}H_{0}^{2} +m \bigg(\Psi^{2 + 2
q}H_{0}^{2}m^{-2}\bigg)^{-3\frac{m}{8}}\bigg\}
\\\nonumber
&+\Psi^{2 + 2 q} H_{0}^{2}\bigg\{72 (\Psi)^{2 + 2 q} H_{0}^{2}
-\sqrt{(\Psi)^{2 + 2 q} H_{0}^{2}}+3m\bigg(\frac{(\Psi)^{2 + 2 q}
H_{0}^{2}}{m^{2}}\bigg)^{-3\frac{m}{8}}\bigg\}\bigg\}\bigg\}\bigg\}
\\\nonumber
&\times\bigg[4(4 + 3 m) \beta^{3}\bigg\{{\frac{\rho_{o}}{a_{o}^{3}}}
+ \frac{1}{2}\bigg\{-1 - 12(\Psi)^{2 + 2 q}H_{0}^{2} + 2\sqrt{m}
\alpha((\Psi)^{2 + 2 q}H_{0}^{2})^{\frac{1}{4}}
\\\nonumber
&-4\bigg(\frac{(\Psi)^{2 + 2 q}
H_{0}^{2}}{m^{2}}\bigg)^{-3\frac{m}{8}}\bigg\}\bigg\}^{2}\bigg]^{-1}\bigg],
\\\nonumber
s&= \bigg[-4(4 +3 m)(\Psi)^{2+ 2
q}H_{0}^{2}\bigg\{\frac{\rho_{0}}{a_{0}^{3}}
+\frac{1}{2}\bigg\{-1-4\bigg(\frac{(\Psi)^{2+2q}
H_{0}^{2}}{m^{2}}\bigg)^{\frac{-3m}{8}}-12(\Psi)^{2+2q}
\\\nonumber
&\times H_{0}^{2} +2 \sqrt{m}\alpha\big((\Psi)^{2 + 2
q}H_{0}^{2}\big)^{\frac{1}{4}}\bigg\}\bigg\}-\bigg\{\frac{1}{4}
\times\bigg\{2-\bigg\{\Psi^{-2(1+q)}\bigg\{-3m^{\frac{5}{2}}
\alpha\big(\Psi^{2+2q}H_{0}^{2}\big)^{\frac{1}{4}}
\\\nonumber
&-9m^{3}\bigg(\frac{\Psi^{2
+2q}H_{0}^{2}}{m^{2}}\bigg)^{\frac{-3m}{8}}\big(\Psi^{2+2q}H_{0}^{2}\big)^{\frac{5}{4}}
+m^{\frac{3}{2}}\alpha
\big((\Psi)^{2+2q}H_{0}^{2}\big)^{\frac{1}{4}}(-4+54\Psi^{2+2q}
H_{0}^{2})
\\\nonumber
&+12\Psi^{2+2q}H_{0}^{2} \bigg\{\sqrt{\frac{\rho_{0}}{a_{0}^{3}}}
+36\Psi^{2+2q}H_{0}^{2}-2\sqrt{6}\sqrt{\Psi^{2+2q}H_{0}^{2}} -4
\bigg(\frac{\Psi^{2+2q}H_{0}^{2}}{m^{2}}\bigg)^{\frac{-3m}{8}}\bigg\}
+12
\\\nonumber
&\times
m^{2}\bigg\{6\Psi^{2+2q}H_{0}^{2}-\bigg(\frac{(\Psi)^{2+2q}H_{0}^{2}}{m^{2}}\bigg)^{\frac{-3m}{8}}+m\Psi^{2+2q}
H_{0}^{2}\bigg\{9\sqrt{\frac{\rho_{0}}{a_{0}^{3}}}+324 \Psi^{2+2q}
H_{0}^{2}-18
\\\nonumber
&\times\sqrt{6}\sqrt{\Psi^{2+2q}H_{0}^{2}}+
36\bigg(\frac{\Psi^{2+2q}H_{0}^{2}}{m^{2}}\bigg)^{\frac{-3m}{8}}\bigg\}\bigg\}\bigg\}
\bigg[(4+3m)H_{0}^{2}\bigg\{\frac{\rho_{0}}{a_{0}^{3}}
+\frac{1}{2}\bigg\{-\big(\Psi^{2+2q}\beta)^{2}
\\\nonumber
&+2\sqrt{m}\alpha\big(\Psi)^{2+2q} \beta)^{2}\big)^{\frac{1}{4}}-4
\bigg(\frac{((\Psi)^{2+2q}
\beta)^{2}}{m^{2}}\bigg)^{\frac{-3m}{8}}\bigg\}\bigg\}\bigg]^{-1}
+8\bigg\{\frac{1}{2}-\bigg\{\Psi^{1+q)}\bigg\{m^{\frac{5}{2}}
\\\nonumber &
\times\alpha\big(\Psi^{2+2q}H_{0}^{2}\big)^{\frac{1}{4}}
+72\sqrt{m}\alpha(\Psi^{2+2q}H_{0}^{2})^{\frac{5}{4}})
-9m^{3}\bigg(\frac{\Psi^{2+2q}H_{0}^{2}}{m^{2}}\bigg)^{-3
\frac{m}{8}}+\big(\Psi^{2+2q} H_{0}^{2}\big)^{\frac{1}{4}}
\\\nonumber
&+m^{\frac{3}{2}}\alpha\bigg\{54(\Psi)^{2+2q}H_{0}^{2}
+12\Psi^{2+2q}H_{0}^{2}\bigg\{-1-\sqrt{\frac{\rho_{0}}{a_{0}^{3}}}
+36\Psi^{2 + 2 q}H_{0}^{2}-2\sqrt{6}
\\\nonumber
&\times\sqrt{\Psi^{2+2q}H_{0}^{2}}-4\bigg(\frac{\Psi)^{2+2q}
H_{0}^{2}}{m^{2}}\bigg)^{\frac{-3m}{8}}\bigg\}\bigg\}+
m^{2}\bigg\{6\Psi^{2+2q}H_{0}^{2}
-\bigg(\frac{(\Psi)^{2+2q}H_{0}^{2}}{m^{2}}\bigg)^{\frac{-3m}{8}}\bigg\}
\\\nonumber
&+m\Psi^{2+2q} H_{0}^{2} \bigg\{\sqrt{\frac{\rho_{0}}a_{0}^{3}}+
\Psi^{2+2q} H_{0}^{2}-\sqrt{6}
\sqrt{(\Psi)^{2+2q}H_{0}^{2}}+\bigg(\frac{\Psi^{2+2q}H_{0}^{2}}
{m^{2}}\bigg)^{\frac{-3m}{8}}\bigg\}\bigg\}\bigg\}
\\\nonumber
&\times\bigg[4H_{0}^{2} \bigg\{\frac{\rho_{0}}{a_{0}^{3}}
+\frac{1}{2}\bigg\{ \Psi^{2+2q}H_{0}^{2} +324m^{2}
+2\sqrt{m}\alpha\big(\Psi^{2+2q}H_{0}^{2}\big)^{\frac{1}{4}}-4
\bigg(\frac{\Psi^{2+2q}H_{0}^{2}}{m^{2}}\bigg)^{\frac{-3m}{8}}
\\\nonumber
&+12m^{2} \sqrt{\frac{\rho_{0}}{a_{0}^{3}}} \bigg\}\bigg\}\bigg]^{2}
\bigg[\alpha(\Psi^{2+2q}H_{0}^{2})^{\frac{1}{4}}\bigg\} +\sqrt{m}
\alpha \big(\Psi^{2+2q}H_{0}^{2}\big)^{\frac{5}{4}} -9
m^{3}\bigg(\frac{\Psi^{2+2q}H_{0}^{2}}{m^{2}}\bigg)^{\frac{-3m}{8}}
\\\nonumber
& +m^{\frac{3}{2}}\alpha\big(\Psi
^{2+2q}H_{0}^{2}\big)^{\frac{1}{4}} (-4
+54\Psi^{2+2q}H_{0}^{2})+12\Psi^{2+2q}H_{0}^{2}
\sqrt{\frac{\rho_{0}}{a_{0}^{3}}}+36 \Psi^{2+2q}H_{0}^{2}
\\\nonumber
&-\sqrt{\Psi^{2+2q}H_{0}^{2}}-4
\bigg(\frac{\Psi^{2+2q}H_{0}^{2}}{m^{2}}\bigg)\Psi^{2+2q}
H_{0}^{2}-\bigg(\frac{\Psi
^{2+2q}H_{0}^{2}}{m^{2}}\bigg)^{\frac{-3m}{8}}\bigg\}+m (\Psi)^{2 +
2 q}H_{0}^{2}
\\\label{65}
& - 9\sqrt{\frac{\rho_{0}}{a_{0}^{3}}} +324 \Psi^{2+2q}H_{0}^{2}
-18\sqrt{6}\sqrt{\Psi^{2+2q}H_{0}^{2}} +36
\bigg(\frac{\Psi^{2+2q}H_{0}^{2}}{m^{2}}\bigg)\bigg\}\bigg\}\bigg]^{-1}.
\end{align}

\section*{Appendix F: Determination of $\nu_{s}^{2}$ }
\renewcommand{\theequation}{F\arabic{equation}}
\setcounter{equation}{0}

\begin{align}\nonumber
\nu_{s}^{2}&=\bigg[\Psi^{-2
(1+q)}\bigg\{-4\bigg\{\frac{\rho_{0}}{a_{0}^{3}}+\frac{1}{2}
\bigg\{2\sqrt{m}\alpha (\Psi^{2+2q}H_{0}^{2})^{\frac{1}{4}}
-4\bigg(\frac{\Psi^{2+2q}H_{0}^{2}}{m^{2}} \bigg)^{
\frac{-3m}{8}}\Psi^{2+2q}\bigg\}\bigg\}
\\\nonumber
&\times H_{0}^{2}+\bigg\{m^{\frac{5}{2}}\alpha
(\Psi^{2+2q}H_{0}^{2})^{\frac{1}{4}}+
\sqrt{m}\alpha\big(\Psi^{2+2q}H_{0}^{2}\big)^{\frac{5}{4}}
-9m^{3}\bigg(\frac{\Psi^{2+2q}H_{0}^{2}}{m^{2}}\bigg)^{\frac{-3m}{8}}
+m^{\frac{3}{2}}
\\\nonumber
&\times\alpha(\Psi^{2+2q}H_{0}^{2})^{\frac{1}{4}}(-4
+54\Psi^{2+2q}H_{0}^{2}) +12\Psi^{2+2q}H_{0}^{2}\bigg\{-1 -
\sqrt{\frac{\rho_{0}}{a_{0}^{3}}} +36\Psi^{2+2q}H_{0}^{2}
\\\nonumber
&-2\sqrt{6}\sqrt{\Psi^{2+2q}H_{0}^{2}}
-4\bigg(\frac{\Psi^{2+2q}H_{0}^{2}}{m^{2}}\bigg)\bigg\}+\bigg\{\Psi^{2+2q}
H_{0}^{2}-\bigg(\frac{\Psi^{2+2q}H_{0}^{2}}{m^{2}}\bigg)\Psi^{
2+2q}H_{0}^{2}\\\nonumber
&\times\bigg\{87-9\sqrt{\frac{\rho_{0}}{a_{0}^{3}}}+324
\Psi^{2+2q}H_{0}^{2}-\sqrt{\Psi^{2+2q}H_{0}^{2}}+36
\bigg(\frac{\Psi^{2+2q}H_{0}^{2}}{m^{2}}\bigg)+\bigg\{\bigg(
\Psi^{2+2q} \\\nonumber &\times
H_{0}^{2}+2\sqrt{m}\alpha(\Psi^{2+2q}H_{0}^{2})^{\frac{1}{4}}\bigg)-4
\bigg(\frac{\Psi^{2+2q}H_{0}^{2}}{m^{2}}\bigg)^{\frac{-3m}{8}}\bigg\}\bigg\}
\bigg\{2(-24\Psi^{2+2q}H_{0}^{2} +\sqrt{m}
\\\nonumber &\times\alpha(\Psi^{2+2q}H_{0}^{2})^{\frac{1}{4}})+3 m
\bigg(\frac{\Psi^{2+2q}H_{0}^{2}}{m^{2}}\bigg)^{\frac{-3m}{8}}
-\bigg\{\alpha(\Psi^{2+2q}H_{0}^{2})^{\frac{1}{4}})+
\sqrt{m}\alpha(\Psi^{2+2q}H_{0}^{2})^{\frac{5}{4}} \\\nonumber
&-m^{3}
\bigg(\frac{\Psi^{2+2q}H_{0}^{2}}{m^{2}}\bigg)^{\frac{-3m}{8}}
+m^{\frac{3}{2}}
\alpha\big(\Psi^{2+2q}H_{0}^{2}\big)^{\frac{1}{4}}(-4+54
\Psi^{2+2q}H_{0}^{2})+12\Psi^{2+2q}
\\\nonumber
&\times H_{0}^{2}\bigg\{2m^{2}-
\sqrt{\frac{\rho_{0}}{a_{0}^{3}}}+36\Psi^{2+2q}H_{0}^{2}-2\sqrt{6}
\sqrt{\Psi^{2+2q}H_{0}^{2}} -4
\bigg(\frac{\Psi^{2+2q}H_{0}^{2}}{m^{2}}\bigg)^{\frac{-3m}{8}}
+m^{2}
\\\nonumber &\times\bigg\{\Psi^{2+2q}
H_{0}^{2}-\bigg(\frac{\Psi^{2+2q}H_{0}^{2}}
{m^{2}}\bigg)^{\frac{-3m}{8}}\bigg\}\bigg\}+m\Psi^{
2+2q}H_{0}^{2}\bigg\{\sqrt{m}\alpha
-\sqrt{\frac{\rho_{0}}{a_{0}^{3}}} +324 \Psi^{2+2q}
\\\nonumber
&\times H_{0}^{2}-18\sqrt{6}\sqrt{\Psi^{2+2q}H_{0}^{2}}+36
\bigg(\frac{\Psi^{2+2q}H_{0}^{2}}{m^{2}}\bigg)+
\bigg\{\bigg\{\Psi^{2+2q}H_{0}^{2}
+\alpha(\Psi^{2+2q}H_{0}^{2})^{\frac{1}{4}}\bigg\}
\\\nonumber & -\bigg(\frac{\Psi^{2+2q}H_{0}^{2}}{m^{2}}\bigg)^{\frac{-3m}{8}}\bigg\}\bigg\}
-3m^{\frac{5}{2}}
\alpha\big(\Psi^{2+2q}H_{0}^{2}\big)^{\frac{1}{4}}+
(\Psi^{2+2q}H_{0}^{2})^{\frac{5}{4}}-
\bigg(\frac{\Psi^{2+2q}H_{0}^{2}}{m^{2}}\bigg)^{\frac{-3m}{8}}
\\\nonumber
&\times\sqrt{m}\alpha+m^{\frac{3}{2}}\alpha(\Psi^{2+2q}H_{0}^{2})^{\frac{1}{4}}(-4
+ 54\Psi^{2+2q}H_{0}^{2})+12\Psi ^{2+2q}H_{0}^{2}+324m^{2}
\\\nonumber
&-1-\sqrt{\frac{\rho_{0}}{a_{0}^{3}}}
+\Psi^{2+2q}H_{0}^{2}-2\sqrt{6} \sqrt{\Psi^{2+2q}H_{0}^{2}}-4
\bigg(\frac{\Psi^{2+2q}H_{0}^{2}}{m^{2}}\bigg)^{2}+\bigg\{6\Psi^{2+
2q}H_{0}^{2}\\\nonumber &-\bigg(\frac{\Psi^{2+2q}H_{0}^{2}}{
m^{2}}\bigg)^{\frac{-3m}{8}}\bigg\}+m\Psi^{2+2q}H_{0}^{2}(87 - 9
\sqrt{\frac{\rho_{0}}{a_{0}^{3}}} +324\Psi^{2+2q}H_{0}^{2}
-\sqrt{\Psi^{2+2q}H_{0}^{2}}
\\\nonumber
&\times18\sqrt{6}+36\sqrt{m}\bigg(\frac{\Psi^{2+2q}H_{0}^{2}}{m^{2}}\bigg)^{\frac{-3m}{8}}
-\bigg\{\frac{\rho_{0}}{a_{0}^{3}} +\frac{1}{2}\bigg\{-1
+2\sqrt{m}\alpha(\Psi^{2+2q}H_{0}^{2})^{\frac{1}{4}}-12
\\\nonumber
&\times\Psi^{2+2q}H_{0}^{2}\bigg\}+12m^{2}-4
\bigg(\frac{\Psi^{2+2q}H_{0}^{2}}{m^{2}}\bigg)^{\frac{-3m}{8}}\bigg\}
-m^{\frac{5}{2}}\alpha
\big(\Psi^{2+2q}H_{0}^{2}\big)^{\frac{1}{4}})+
\big(\Psi^{2+2q}\big)^{\frac{5}{4}}
\\\nonumber
&\times\sqrt{m} \alpha+m^{\frac{3}{2}}\alpha
\big(\Psi^{2+2q}H_{0}^{2}\big)^{\frac{5}{4}}+m^{4}
\bigg(\frac{\Psi^{2+2q}H_{0}^{2}}{m^{2}}\bigg)^{\frac{-3m}{8}}+m^{\frac{3}{2}}
\alpha(\Psi^{2+2q}H_{0}^{2})^{\frac{1}{4}}-2
\\\nonumber
&+27\Psi^{2+2q}H_{0}^{2}+96(\Psi)^{2+2q}H_{0}^{2}(-1 -
\sqrt{\frac{\rho_{0}}{a_{0}^{3}}}+36\alpha\Psi^{2+2q}H_{0}^{2}
-2\sqrt{6} \sqrt{\Psi^{2+2q}H_{0}^{2}} \\\nonumber &-4
\bigg(\frac{\Psi^{2+2q}H_{0}^{2}}{m^{2}}\bigg)\Psi^{2+2q}
H_{0}^{2}(87-9\sqrt{\alpha}\sqrt{\frac{\rho_{0}}{a_{0}^{3}}}+324
\Psi^{2+2q}H_{0}^{2}-\sqrt{\Psi^{2+2q}H_{0}^{2}} \\\nonumber &+36
\bigg(\frac{\Psi^{2+2q}H_{0}^{2}}{m^{2}}\bigg)\Psi^{2 + 2 q}
\bigg\{72 \Psi^{2+2q}H_{0}^{2}-2\sqrt{6}
\sqrt{\Psi^{2+2q}H_{0}^{2}}-3m
\bigg(\frac{\Psi^{2+2q}H_{0}^{2}}{m^{2}}\bigg)
\\\nonumber
&\times36m^{2}\bigg\{16 \Psi^{2+2q}H_{0}^{2}+m
\bigg(\frac{\Psi^{2+2q}H_{0}^{2}}{m^{2}}\bigg)\Psi^{2+2q}
H_{0}^{2}\bigg\{72 \Psi^{2+2q}H_{0}^{2}-2\sqrt{6}
\\\nonumber
&\times\sqrt{\Psi^{2+2q}H_{0}^{2}}
+3m\bigg(\frac{\Psi^{2+2q}H_{0}^{2}}{m^{2}}\bigg)^{\frac{-3m}{8}}
\bigg\}\bigg\}\bigg\}\bigg[-\Psi^{2+2q}H_{0}^{2}
+\sqrt{m}\alpha(\Psi^{2+2q}H_{0}^{2})^{\frac{1}{4}}
\\\nonumber
&+3m\bigg(\frac{\Psi^{2+2q}H_{0}^{2}}{m^{2}}\bigg)\bigg]\bigg[(24 (4
+ 3 m) H_{0}^{2}\bigg\{\frac{\rho_{0}}{a_{0}^{3}}+\frac{1}{2}
\bigg\{\Psi^{2+2q}H_{0}^{2} +2\sqrt{m}\alpha
(\Psi^{2+2q}H_{0}^{2})^{\frac{1}{4}}
\\
&-4
\bigg(\frac{\Psi^{2+2q}H_{0}^{2}}{m^{2}}\bigg)^{2}\bigg\}\bigg\}^{2}\bigg]^{-1}.
\end{align}
\\
\textbf{Data availability:} No new data were generated or analyzed
in support of this research.

\end{document}